\documentclass[usenatbib]{mn2e}
\usepackage{graphicx}
\usepackage{amssymb}
\usepackage{amsmath}
\usepackage{journal_names}
\usepackage{breqn}

\title[SAMI shapes]{The SAMI Galaxy Survey: the intrinsic shape of kinematically selected galaxies}
\author[C. Foster et al.]{C. Foster,$^{1}$\thanks{E-mail: cfoster@aao.gov.au} J. van de Sande,$^{2}$ F. D'Eugenio$^{3}$, L. Cortese,$^{4}$ R. M. McDermid,$^{1,5}$\newauthor J. Bland-Hawthorn,$^{2,6}$ S. Brough,$^7$ J. Bryant,$^{1,2,6}$ S. M. Croom,$^{2,6}$ M. Goodwin,$^{1}$\newauthor I. S. Konstantopoulos,$^1$ J. Lawrence,$^1$  \'A. R. L\'opez-S\'anchez,$^{1,5}$ A. M. Medling,$^{3,8,9}$\newauthor M. S. Owers,$^{1,5}$ S. N. Richards,$^{1,2,7}$  N. Scott,$^2$ D. S. Taranu,$^{4,7}$ C. Tonini,$^{10}$ and \newauthor T. Zafar$^1$
\\
$^1$Australian Astronomical Observatory, 105 Delhi Rd, North Ryde, NSW 2113, Australia\\
$^2$Sydney Institute for Astronomy, School of Physics, A28, The University of Sydney, NSW, 2006, Australia\\
$^3$Research School of Astronomy and Astrophysics, Australian National University, Canberra ACT 2611, Australia\\
$^4$International Centre for Radio Astronomy Research, University of Western Australia, 35 Stirling Highway, Crawley WA 6009, Australia\\
$^5$Department of Physics and Astronomy Macquarie University NSW 2109 Australia\\
$^6$ARC Centre of Excellence for All-sky Astrophysics (CAASTRO), The University of Sydney, NSW 2006, Australia\\
$^7$School of Physics, University of New South Wales, NSW 2052, Australia\\
$^8$Cahill Center for Astronomy and Astrophysics California Institute of Technology, MS 249-17 Pasadena, CA 91125, USA\\
$^9$Hubble Fellow\\
$^{10}$Melbourne University, School of Physics Parkville, 3010 Australia
}

\begin{document}

\date{}

\pagerange{\pageref{firstpage}--\pageref{lastpage}} \pubyear{2014}

\maketitle

\label{firstpage}

\begin{abstract}
Using the stellar kinematic maps and ancillary imaging data from the Sydney AAO Multi Integral field (SAMI) Galaxy Survey, the intrinsic shape of kinematically-selected samples of galaxies is inferred. We implement an efficient and optimised algorithm to fit the intrinsic shape of galaxies using an established method to simultaneously invert the distributions of apparent ellipticities and kinematic misalignments. The algorithm output compares favourably with previous studies of the intrinsic shape of galaxies based on imaging alone and our re-analysis of the ATLAS$^{\rm 3D}$ data. Our results indicate that most galaxies are oblate axisymmetric. We show empirically that the intrinsic shape of galaxies varies as a function of their rotational support as measured by the ``spin'' parameter proxy $\lambda_{R_e}$. In particular, low spin systems have a higher occurrence of triaxiality, while high spin systems are more intrinsically flattened and axisymmetric. The intrinsic shape of galaxies is linked to their formation and merger histories. Galaxies with high spin values have intrinsic shapes consistent with dissipational minor mergers, while the intrinsic shape of low-spin systems is consistent with dissipationless multi-merger assembly histories. This range in assembly histories inferred from intrinsic shapes is broadly consistent with expectations from cosmological simulations.

\end{abstract}


\section{Introduction}

The true or intrinsic shape of galaxies in three-dimensions is a difficult property to measure due to projection effects. Many studies have attempted to measure this fundamental characteristic of galaxies over the last 90 years (\citealt{Hubble26}; also see \citealt{MendezAbreu16} and references therein for a recent review). However, a full understanding of the distribution of intrinsic shapes of galaxies and its dependence on other galaxy properties is only just starting to be uncovered.

Theoretical simulations suggest that the process of galaxy formation influences the intrinsic shape of galaxies. For example, gas dissipation in galaxy mergers leads to central star formation, which alters the gravitational potential and changes the intrinsic shape of the remnant \citep{Cox06}. Galaxies that were formed through dissipative processes tend to be more flattened. The merger histories of galaxies also influence their intrinsic shape. Theoretical simulations show that both the merger mass ratio \citep{Jesseit09} and frequency of mergers \citep{Moody14} alter the intrinsic shape of galaxies in predictable ways. Galaxies that experience major mergers (i.e. high progenitor mass ratios) and/or more frequent mergers tend to be more triaxial \citep[e.g.][]{Jesseit09,Taranu13}. Discs of galaxies in triaxial dark matter haloes are predicted to be slightly elliptical \citep{Bailin07}. In other words, measuring the intrinsic shape of real galaxies can give new insights into their formation, merger and star formation histories.

The advent of large photometric surveys such as the Two Micron All-Sky Survey \citep[2MASS;][]{Jarrett03} and the Sloan Digital Sky Survey \citep[SDSS;][]{York00} have allowed for the inversion of the distribution of apparent axis ratios into a parent distribution of intrinsic axis ratios through a set of reasonable assumptions \citep[e.g.][]{Ryden06,Padilla08,Rodriguez13}. While the distribution of apparent axis ratios of galaxies can constrain the intrinsic flattening, it is essentially insensitive to the circularity of the disc, hence leaving little handle on triaxiality. Such studies have estimated the overall intrinsic shape of galaxies \citep{Kimm07}, and identified variations in the intrinsic shape of galaxies as a function of environment \citep[e.g.][]{Ryden93,Fasano10,Rodriguez16}, luminosity \citep[e.g.][]{SanchezJanssen16}, stellar mass \citep[e.g.][]{SanchezJanssen10, Holden12}, redshift \citep[e.g.][]{Holden12} and morphology \citep[e.g.][]{Ryden06,Padilla08,Rodriguez13}. 

The added dimension obtained through kinematic mapping is required to reliably constrain the intrinsic shape of individual elliptical galaxies \citep[e.g.][]{Statler94a, vandenBosch09}, but these models have degeneracies between triaxiality and inclination angle. For statistical samples of galaxies, the method of \citet[][]{Franx91} offers the best handle on triaxiality because large samples allow for marginalisation over all possible inclination angles. In recent years, this method of inverting the projected shape and kinematic misalignment distributions proposed by \citet{Franx91} has been successfully implemented by the ATLAS$^{\rm 3D}$ \citep{Weijmans14} and the SAGES Legacy Unifying Globulars and GalaxieS \citep[SLUGGS;][]{Foster16} survey teams with the availability of stellar kinematic maps for sufficiently large samples. However, the parent sample size of these surveys (260 targets or less) was insufficient to probe variations in the intrinsic shape of galaxies with other fundamental properties.

The development of multiplexed integral field spectrographs (multi-IFS, e.g. \citealt{Croom12}) such as the SAMI instrument in the last decade now enables surveys of thousands of galaxies to be carried out. The SAMI Galaxy Survey \citep{Bryant15,Green17} has already mapped over 1500 galaxies using multi-IFS technology and upon completion will have gathered over 3000 galaxy spectral cubes. In the northern hemisphere, the SDSS-IV MaNGA Survey (Sloan Digital Sky Survey Data; Mapping Nearby Galaxies at APO; \citealt{Bundy15}) is also obtaining 3D spectroscopy of thousands of galaxies. These data have the potential to revolutionise our understanding of galaxy intrinsic shapes and their dependence on other fundamental properties. These new insights into the shape of galaxies will enable detailed comparisons with theoretical simulations of galaxy formation.

In this work, we study the intrinsic shape of kinematically selected galaxies in the SAMI Galaxy Survey. The paper is divided as follows: our data and sample selection are presented in Section \ref{sec:data}, while Section \ref{sec:method} presents the method we use to infer the intrinsic shape of galaxies. Our analysis and results are presented in Section \ref{sec:analysis}. Section \ref{sec:discussion} contains a discussion of our results, while a summary and our conclusions are outlined in Section \ref{sec:conclusion}.

We assume a $\Lambda$CDM cosmology with  $\Omega_{\rm m}=0.3$, $\Omega_{\lambda}=0.7$ and $H_0=70$ km s$^{-1}$ Mpc$^{-1}$.

\section{Data}\label{sec:data}

\begin{table*}
\begin{center}
\begin{tabular}{ccccccccc}
\hline
Sample & Selection & $N_{\rm gals}$ & $\mu_Y$ & $\sigma_Y$ & $\mu_q$ & $\sigma_q$ & $A^2$ & $N_{\rm it}$ \\
(1)&(2)&(3)&(4)&(5)&(6)&(7)&(8)&(9)\\
\hline 
Fast Rotators & $\lambda_{R_e}\ge0.31\sqrt{\epsilon_{R_e}}$ & 766 & -5.08 & 1.95 & 0.32 & 0.24 & 0.0005 & 4 \\
Slow Rotators & $\lambda_{R_e}<0.31\sqrt{\epsilon_{R_e}}$ & 79 & -0.38& 3.07 & 0.62 & 0.08 & 0.0009 & 200 \\
R1 & $\lambda_{R_e}<0.2$ & 187 & -1.27 & 3.37 & 0.85 & 0.30 & 0.0004 & 2 \\
R2 & $0.2\le\lambda_{R_e}<0.3$ & 178 & -4.16 & 2.46 & 0.44 & 0.05 & 0.0004 & 19 \\
R3 & $0.3\le\lambda_{R_e}<0.45$  & 238 & -5.39 & 1.58 & 0.36 & 0.05 & 0.0004 & 8 \\
R4 & $0.45\le\lambda_{R_e}<0.75$  & 242 & -5.40 & 1.29 & 0.27 & 0.04 & 0.0012 & 200 \\ 
\hline
\end{tabular}\caption{Summary of the intrinsic shape outputs for our selected sub-samples (Column 1) assuming the intrinsic kinematic misalignment follows $\tan\Psi_{\rm int}=\sqrt{T/(1-T)}$. The selection criteria for each sample are listed in Column 2. The number of galaxies in each sample is listed in Column 3. The mean $\mu_Y$ and standard deviation $\sigma_Y$ of the lognormal distribution in the intermediate over long axis ratio ($p$ such that $Y=1-\ln(p)$) are given in Columns 4 and 5, respectively. Columns 6 and 7 report the mean $\mu_q$ and standard deviation $\sigma_q$ of the short over long axis ratio ($q$). The figure of merit $A^2$ is given in Column 8 for each sample with $A^2=0$ equivalent to a perfect fit. Column 9 lists the number of iterations performed before achieving convergence (i.e. $A^2\le0.0005$) with the maximum set at 200 iterations.}
\label{table:shapesoutput}
\end{center}
\end{table*}

\begin{figure*}
\begin{center}
\includegraphics[width=175mm]{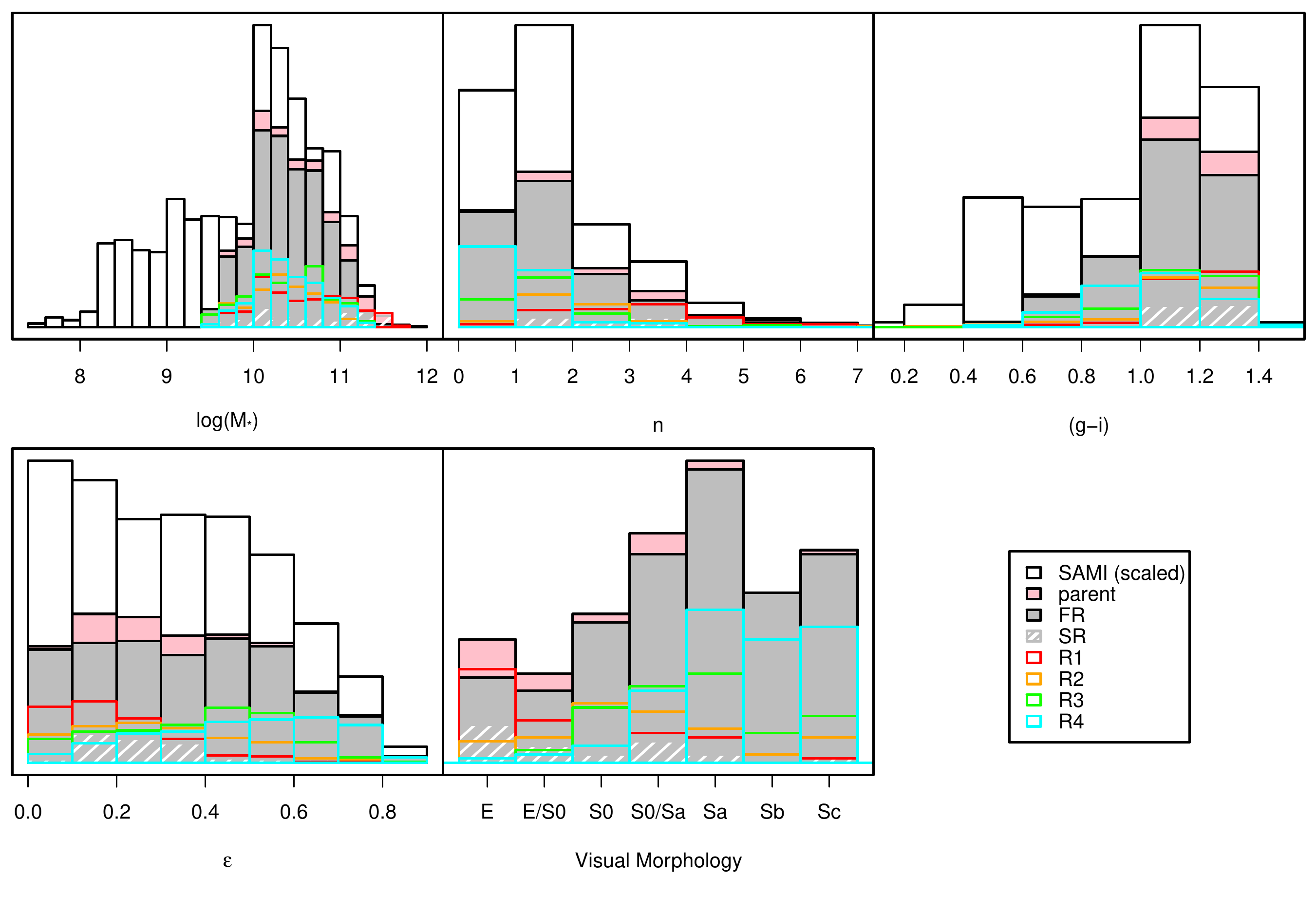}
\caption{Histograms summarising the distribution of stellar mass in $\log(M_*[{\rm M_\odot}])$ (top left), S\'ersic index (n, top middle) for the GAMA sample only (i.e. no cluster galaxies), $(g-i)$ colours (top right), global ellipticities ($\epsilon$, lower left) and visual morphologies as per \citet{Cortese16} (lower middle). The distributions are shown for the SAMI primary target sample (white \emph{scaled} histograms), the `parent' sample (pink), fast rotators (grey with black border), slow rotators (grey with white hash), R1 (red), R2 (orange), R3 (green) and R4 (cyan). Visual morphologies are currently only available for a subset of the observed SAMI galaxies and are not necessarily representative, especially for small samples.}\label{fig:sample_properties}
\end{center}
\end{figure*}

\begin{figure}
\begin{center}
\includegraphics[width=84mm]{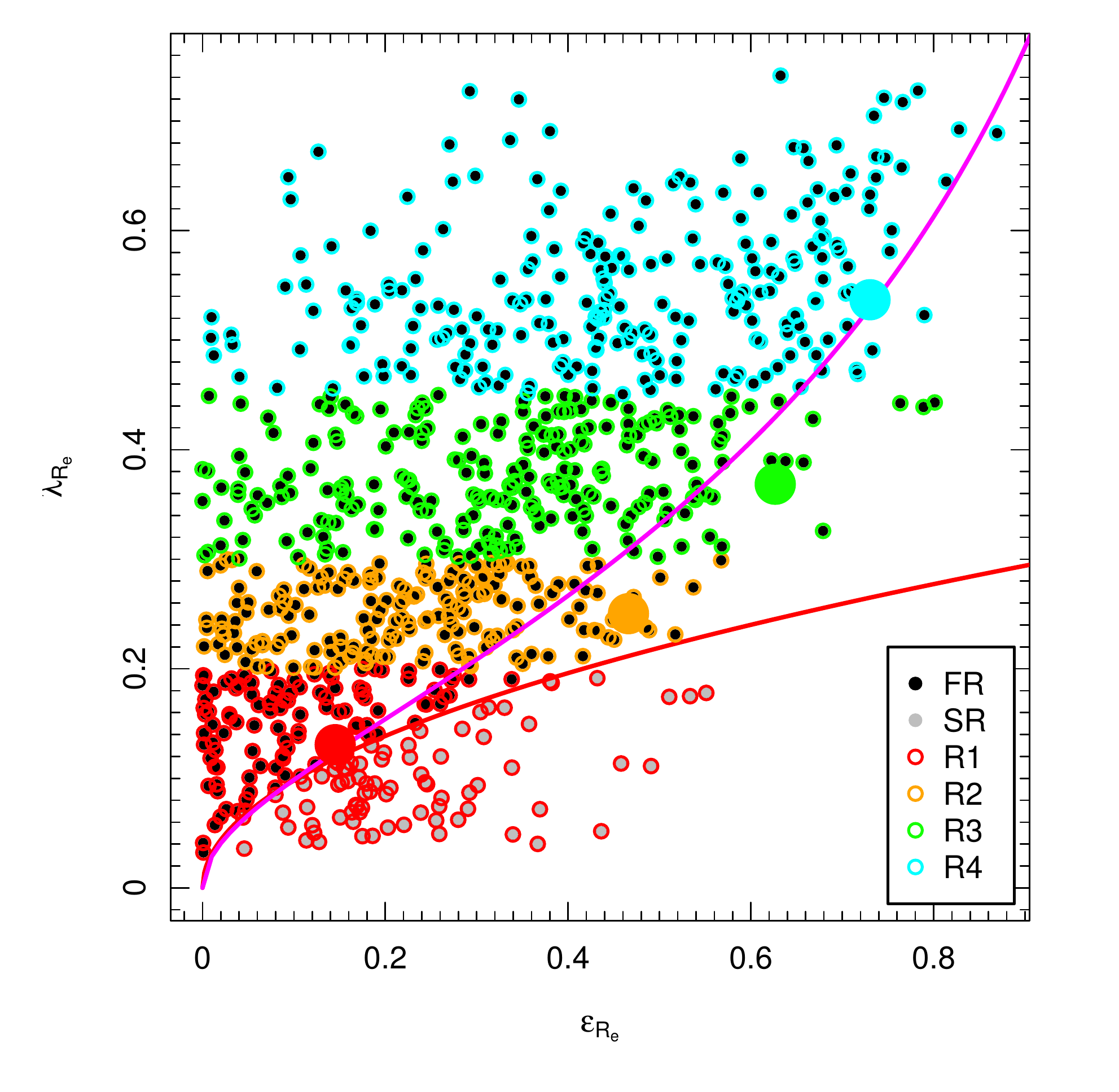}
\caption{Spin parameter proxy ($\lambda_{R_e}$) versus ellipticity measured at $R_e$ ($\epsilon_{R_e}$) for our parent sample. Red solid line shows the separation between fast (black points) and slow (grey points) rotators as per \citet{Emsellem11}. The parent sample is divided into four subsamples of increasing rotational support (R1, red circles; R2, orange circles; R3, green circles and R4, cyan circles) are separated using lines of constant $\lambda_{R_e}=0.2, 0.3, 0.45, 0.7$, see text. Large points of corresponding colours show the median $\lambda_{R_e}$ and intrinsic (edge-on) ellipticity for each sample. The magenta line shows the semi-empirical prediction for edge-on axisymmetric galaxies with anisotropy parameter $\beta=0.70 \epsilon_{\rm intr}$ \citep[e.g.][]{Cappellari07,Cappellari16a,vandeSande17a}.}\label{fig:FRSR_selection}
\end{center}
\end{figure}

The SAMI spectrograph \citep{Croom12} is a multiplexed instrument capable of obtaining 3D spectroscopy of 12 galaxies and one calibrator star simultaneously using hexabundles \citep{BlandHawthorn11,Bryant14}. Each hexabundle of 61 fibres has a high filling factor (73\%) and diameter of 15 arcsec. The 13 hexabundles and 26 individual sky fibres are fed into the the AAOmega spectrograph \citep{Sharp06} on the 3.9m Anglo-Australian Telescope, which provides a 1 degree field of view. The SAMI Galaxy Survey has so far observed roughly half of the targeted 3600 galaxies between redshifts $z=0.004$ and $z=0.095$. Sampled galaxies cover a broad range of environments, stellar masses, structural properties, colours and morphological types (see Fig. \ref{fig:sample_properties}). The SAMI target selection is described in detail in \citet{Bryant15} for the Galaxy And Mass Assembly (GAMA, \citealt{Driver11}) fields and \citet{Owers17} for the clusters sample. 

The AAOmega setup for the SAMI Galaxy Survey uses the 580V and 1000R gratings, which yields a generous wavelength coverage of 3700-5700 \AA\ and 6300-7400\AA\ with spectral resolution of $R\sim1810$ and $R\sim4260$ in the red and blue arms, respectively. Survey data are reduced using the {\sc 2dfdr} pipeline, which applies bias frames, flat fielding, cosmic ray removal, wavelength calibration using CuAr arc frames, sky subtraction using dedicated sky fibres and spectrum extraction for each fibre using the technique outlined in \citet{Sharp10}. \citet{Hopkins13} gives a thorough summary of the spectral data reduction steps performed within {\sc 2dfdr}. Flux calibration is then applied using the primary standard star observed on the same night as the observations. The flux is scaled and telluric absorption is corrected using the secondary standard star that is observed simultaneously for each field. Because the SAMI hexabundles have a fill-factor of 73\%, galaxies are observed using a dither pattern of typically 7 dithers to ensure continuous spatial sampling is achieved. Datacubes are reconstructed from a minimum of 6 dithers by carefully propagating flux and covariance onto a grid as described in \citet{Sharp15}. Full detail of the SAMI data reduction can be found in \citet{Allen15} and \citet{Sharp15}.

We use the penalised pixel-fitting ({\sc pPXF}, \citealt{Cappellari04, Cappellari16b}) algorithm to parametrise the line-of-sight velocity distribution (LOSVD) for each spaxel using Gauss-Hermite polynomials. The first {\rm two} moments: recession velocity ($V$) and velocity dispersion ($\sigma$) are measured by fitting template spectra to the observed spectra in each spaxel. As described in \citet{vandeSande17a}, spectra within elliptical annuli are first combined into a high signal-to-noise spectrum and fitted in order to determine the best template to be used for fitting the LOSVD of individual member spaxels. This is done to minimise the uncertainties associated with template mismatches on low signal-to-noise spectra. For each spaxel, the corresponding template is convolved with a Gauss-Hermite polynomial representing the LOSVD that minimises the residuals between the observed spectrum and the broadened template. The best-fit LOSVD parameter values ($V$, $\sigma$) for each spaxel are recorded.

Galaxies in the SAMI Galaxy Survey span a range of morphological types, with spectra ranging from star-forming emission line-dominated to quiescent stellar continuum-dominated, depending on the target and/or spaxel location. For this reason, the galaxies do not necessarily contain significant stellar continuum. In this work, we select a parent sample of galaxies with sufficient spatial coverage and stellar continuum signal-to-noise such that reliable velocity ($V$) maps can be extracted, using the least stringent quality cut described in \citet{vandeSande17a}. No morphological, age or galaxy type selection is applied.

Briefly, we select galaxies with velocity and velocity dispersion uncertainties on individual spaxels of $V_{\rm error}<30$ km s$^{-1}$ and $\sigma_{\rm error} < 0.1\sigma + 25$ km s$^{-1}$, respectively, for a minimum filling factor of $>75$\% within the effective radius. Full detail of the quality cut can be found in \citet{vandeSande17a}. We augment our sample by using aperture corrected $\lambda_{R_e}$ values from \citet{vandeSande17b} for galaxies with radial coverage below $R_e$. While this does not alter our conclusions, it increases our sample size by 11 percent. This selection yields a \emph{parent} sample of 845 galaxies to be classified as either fast or slow rotators using a proxy for the spin parameter \citep{Emsellem11}:
\begin{equation}\label{eq:spin}
\lambda_{R}=\frac{\langle R|V| \rangle} {\langle R\sqrt{V^2+\sigma^2} \rangle} = \frac{\Sigma^{N_{\rm spx}}_{i=1} F_iR_i|V_i|}{\Sigma^{N_{\rm spx}}_{i=1} F_iR_i\sqrt{V_i^2+\sigma_i^2}},
\end{equation}
where $R_i$ is the galactocentric radius of the $i^{\rm th}$ spaxel. $F_i$ corresponds to the flux in the $i^{\rm th}$ of $N_{\rm spx}$ spaxel. In this work, and in contrast with \citet{Emsellem11}, we use an elliptical bin to measure $\lambda_{R_e}$ following \citet{vandeSande17a}. Using Equation \ref{eq:spin}, we determine whether $\lambda_{R_e}$ lies above or below $0.31\sqrt{\epsilon_{R_e}}$, where $\epsilon_{R_e}$ is the apparent ellipticity measured at one effective radius, to separate fast and slow rotators, respectively, as per \citet{Emsellem11}. While we nominally chose the \citet{Emsellem11} division for ease of comparison with \citet{Weijmans14}, we also repeated our analysis using the fast and slow rotator division proposed by \citet{Cappellari16a} and this did not significantly alter our conclusions. 

We also subdivide our parent sample into four subsamples of increasing rotational support as defined in Table \ref{table:shapesoutput} and illustrated in Fig. \ref{fig:FRSR_selection}\footnote{We also attempted bounding our fast rotator sub-samples by theoretical lines of constant intrinsic ellipticities $\epsilon_{\rm int}=0.45,0.6,0.75,0.9$ \citep[following][adapted from \citealt{Emsellem07,Cappellari11}]{vandeSande17a} and the semi-empitical prediction for edge-on istropic oblate rotator with anisotropy parameter $\beta=0.7\epsilon_{\rm int}$ \citep{Binney05,Cappellari07}. While this selection should in principle meaningfully divide the fast rotators into families of galaxies with theoretically similar intrinsic shapes, in practice they led to biased multi-modal distributions in ellipticity.}. Fig. \ref{fig:sample_properties} shows that all samples have a similar range and distribution in stellar mass (above the $\log(M_*)=9.5$ threshold) that mimics that of the parent sample, while the distribution of S\'ersic index \citep{Sersic63} shifts progressively to lower values between R1 and R4\footnote{Currently, reliable S\'ersic profiles are only available for the GAMA sample galaxies (i.e. there are no cluster sample galaxies included in the upper middle panel of Fig. \ref{fig:sample_properties}) and this may skew the distributions, especially for small samples.}. We also point out the presence of slow rotators with very low S\'ersic indices. We have visually confirmed that these discy systems exhibit very low central rotation and are likely to be unresolved kinematically decoupled cores and 2$\sigma$ galaxies \citep[see e.g.][]{Krajnovic11}. Galaxies become progressively bluer (lower $(g-i)$ values) from R1 to R4. Each sample shows a range in morphological types with a progressively higher proportion of late-type galaxies from R1 to R4.

\section{Method}\label{sec:method}

Photometric ellipticities and position angles are measured using the Multi-Gaussian Expansion technique \citep[MGE,][]{Emsellem94} and the code from \citet{Cappellari02}. For each galaxy we use a 400 arcsec $r$-band cutout image from the Sloan Digital Sky Survey (SDSS) or VLT Survey Telescope (VST). For Abell 85, where both SDSS and VST images are available, we use the SDSS data. Each input image is processed with {\sc SExtractor} \citep{Bertin11} to derive a mask of neighbouring objects. We use {\sc PSFEx} \citep{Bertin11} to build a model point spread function at the location of the galaxy centre. In order to minimise the effect of asymmetries in the galaxies (e.g. bars, spiral arms) we use the regularised version of MGE, which is designed to reflect the underlying light distribution of each galaxy rather than bars or other non-axisymmetric features \citep{Scott13}. The global ellipticity $\epsilon$ is a light-weighted mean for the whole galaxy, while $\epsilon_{R_e}$ is the ellipticity measured at the effective radius ($R_e$, see D'Eugenio et al. in prep for more detail).

As per \citet{vandeSande17a}, we determine the global kinematic position angle ($PA_{\rm kin}$) using the method described in \citet[][their appendix C]{Krajnovic06}. The mean kinematic misalignment angle ($\Psi$) is then defined as 
\begin{equation}\label{eq:psi}
\sin{\Psi}=|\sin(PA_{\rm phot}-PA_{\rm kin})|,
\end{equation}
following \citet{Franx91}.

\citet{Franx91} have shown that the inclusion of kinematic information (more specifically, measurements of $\Psi$) provides a significant improvement in the determination of intrinsic shapes over inverting the distribution of apparent ellipticities alone. In this work, we invert the distribution of apparent ellipticities ($\epsilon$) and kinematic misalignments ($\Psi$) to obtain the best fitting three dimensional axis ratios describing the intrinsic shape of galaxies in our samples. This is done using the algorithm described in section 3.7 of \citet{Foster16}, which is independent from but based on that described in sections 4.4 and appendix A of \citet{Weijmans14}. 

We give a brief summary of our method in what follows. The intrinsic shape definition is simplified by assuming that galaxies can be approximated as simple ellipsoids with intrinsic axis ratios $p=b/a$ and $q=c/a$ for axes lengths $a \ge b \ge c$ such that $0\le q\le p\le1$. As many galaxies exhibit multiple components (e.g. bulge and disc), the shape parameters derived are for the best ellipsoid equivalent parameters. We need to write the observables as a function of the parameters (i.e. intrinsic shape) that we wish to fit for. The apparent ellipticity $\epsilon$ depends on the intrinsic shape and the line-of-sight projection angles in spherical coordinates $0\le\varphi\le\pi$ and $0\le\nu\le2\pi$ \citep{Contopoulos56}. The measured kinematic misalignment defined in Equation \ref{eq:psi} depends on the intrinsic shape, the intrinsic kinematic misalignment $\Psi_{\rm int}$ (i.e. the angle between the short axis and the rotation axis) and the projection angles. Mathematically, we expect the following dependencies:
\begin{equation}\label{eq:PsiEps}
\Psi=\Psi(\Psi_{\rm int}, p, q, \varphi, \nu);\ {\rm and} \
\epsilon=\epsilon(p,q,  \varphi, \nu).
\end{equation}
As we only have two observables, the problem is under-determined. We therefore have to make some simplifying assumptions, which we describe below.

The observed ellipticity (and eccentricity, $e$) can be re-written as a function of the axis ratios and projection angles as follows (e.g. \citealt{Contopoulos56}):
\begin{equation}
e=(1-\epsilon)^2=\frac{a-\sqrt{b}}{a+\sqrt{b}},
\end{equation}
where
\begin{multline}
a=(1-q^2)\cos^2{\nu}+(1-p^2)\sin^2{\nu}\sin^2{\varphi}+p^2+q^2,\\
b=[(1-q^2)\cos^2{\nu}-(1-p^2)\sin^2{\nu}\sin^2{\varphi}-p^2+q^2]^2+\\
4(1-p^2)(1-q^2)\sin^2{\nu}\cos^2{\nu}\sin^2{\varphi}.
\end{multline}

We define the triaxiality parameter $T$ as per \citet{Franx91}:
\begin{equation}
T=\frac{1-p^2}{1-q^2},
\end{equation}
such that $T=0$ and $T=1$ correspond to perfectly oblate ($a=b$) and prolate systems ($b=c$), respectively, with intermediate values indicating triaxiality ($a\ne b\ne c$). The measured kinematic position angle depends on the line-of-sight angles and the intrinsic misalignment ($\Psi_{\rm int}$).
\begin{equation}\label{eq:pakin}
\tan{(PA_{\rm kin})}=\frac{\sin\varphi\tan\Psi_{\rm int}}{\sin\nu-\cos\varphi\cos\nu\tan\Psi_{\rm int}}
\end{equation}
Following \citet{deZeeuw89}, the observed photometric position angle can be written as a function of the triaxiality parameter and the projection angles via the projection matrix as follows:
\begin{equation}\label{eq:pamin}
\tan(2PA_{\rm min}) =\frac{2T\sin\varphi\cos\varphi\cos\nu}{\sin^2\nu-T(\cos^2\varphi-\sin^2\varphi\cos^2\nu)},
\end{equation}
where $PA_{\rm min}=PA_{\rm phot}+\pi/2$ is the position angle of the projected short axis. Equations \ref{eq:pakin} and \ref{eq:pamin} are substituted into Equation \ref{eq:psi}.

\begin{figure*}
\begin{center}
\includegraphics[width=80mm]{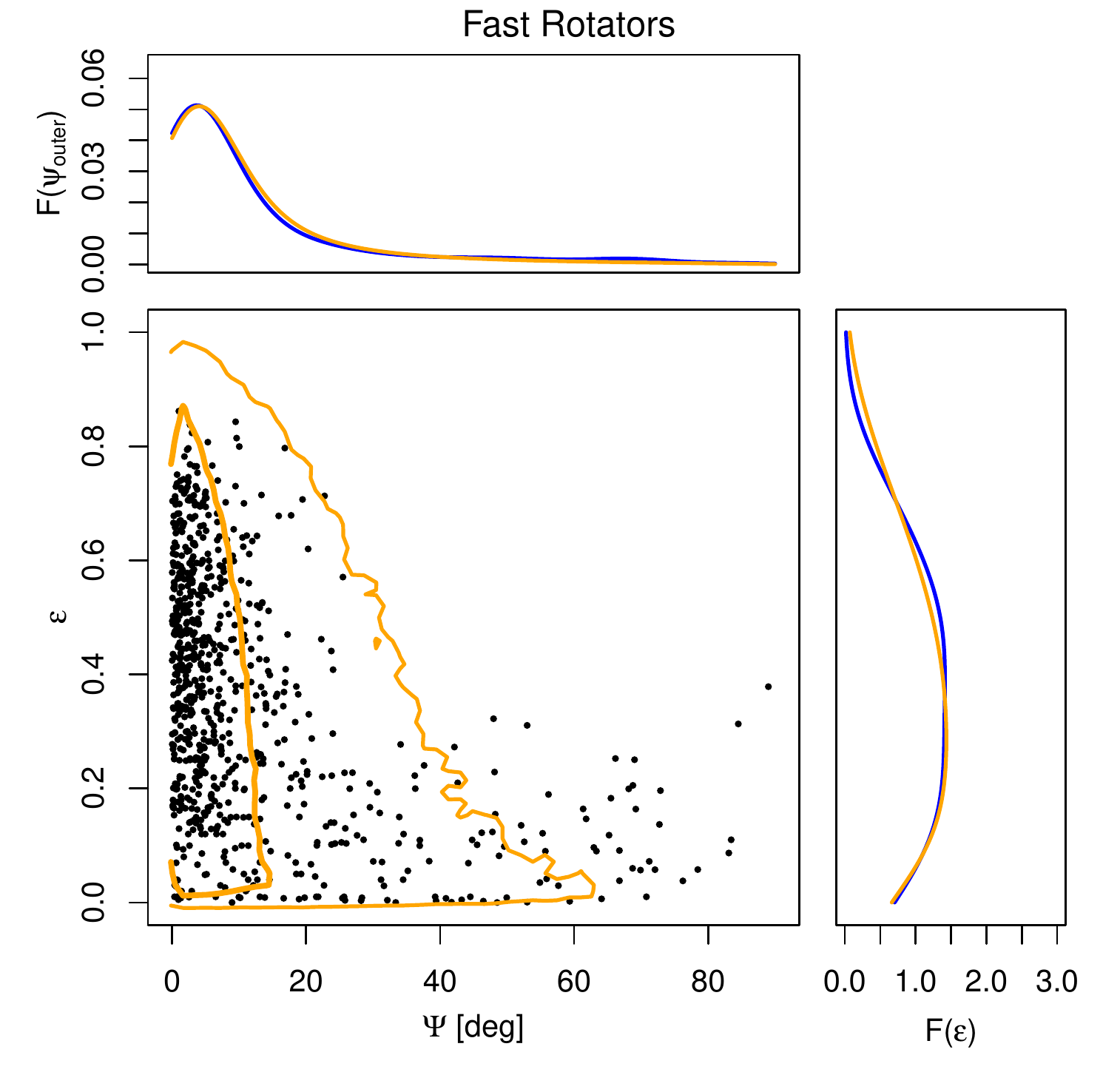}\includegraphics[width=80mm]{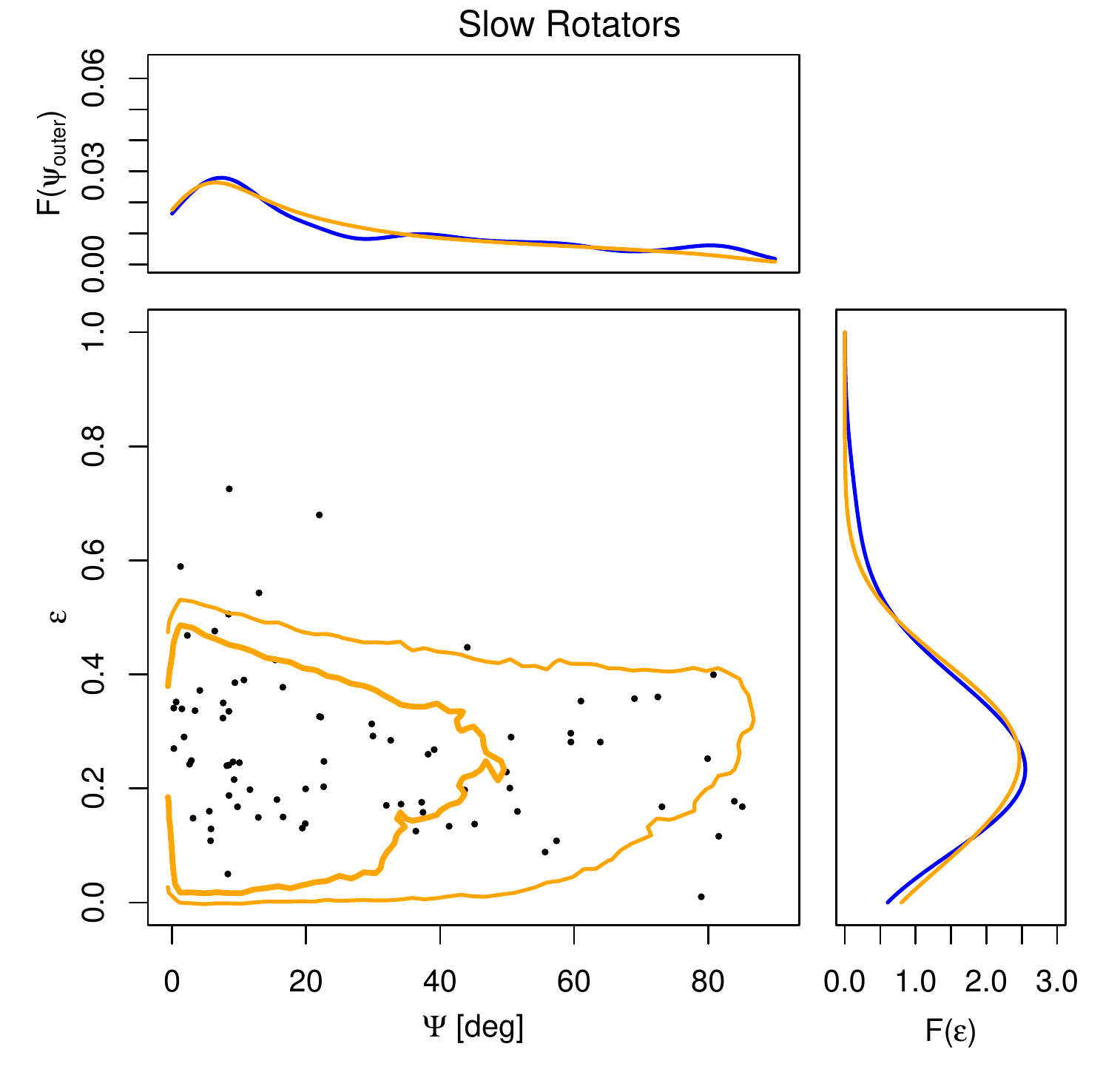}\\
\includegraphics[width=80mm]{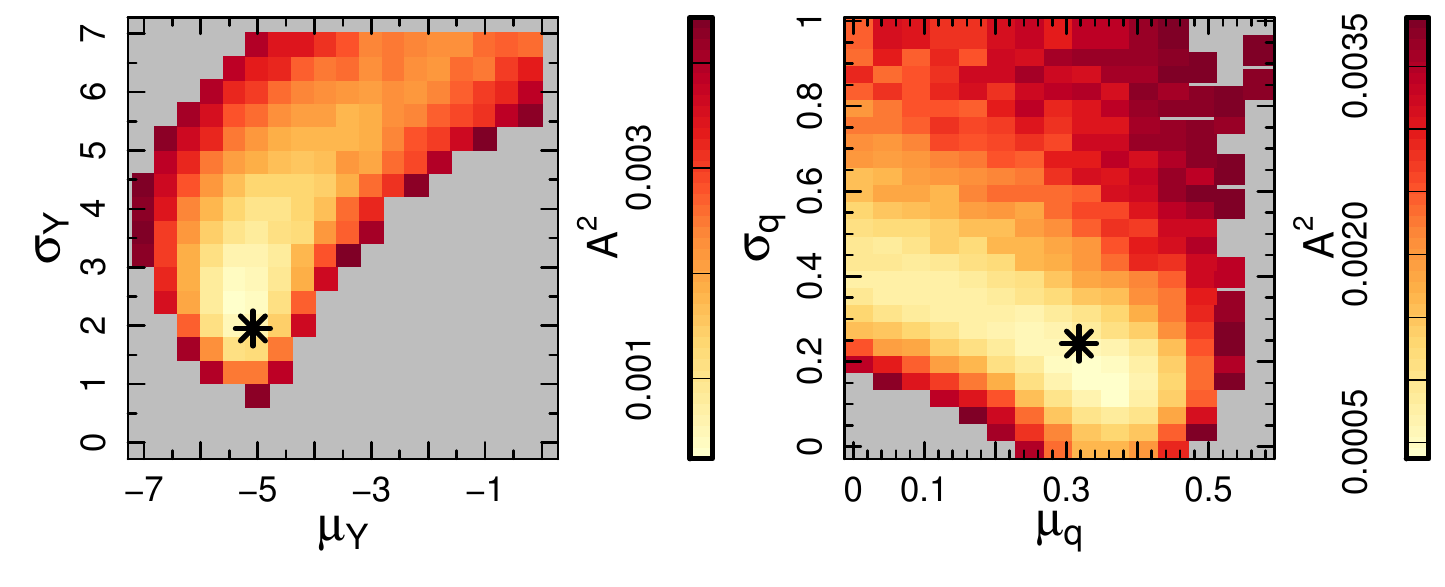}\includegraphics[width=80mm]{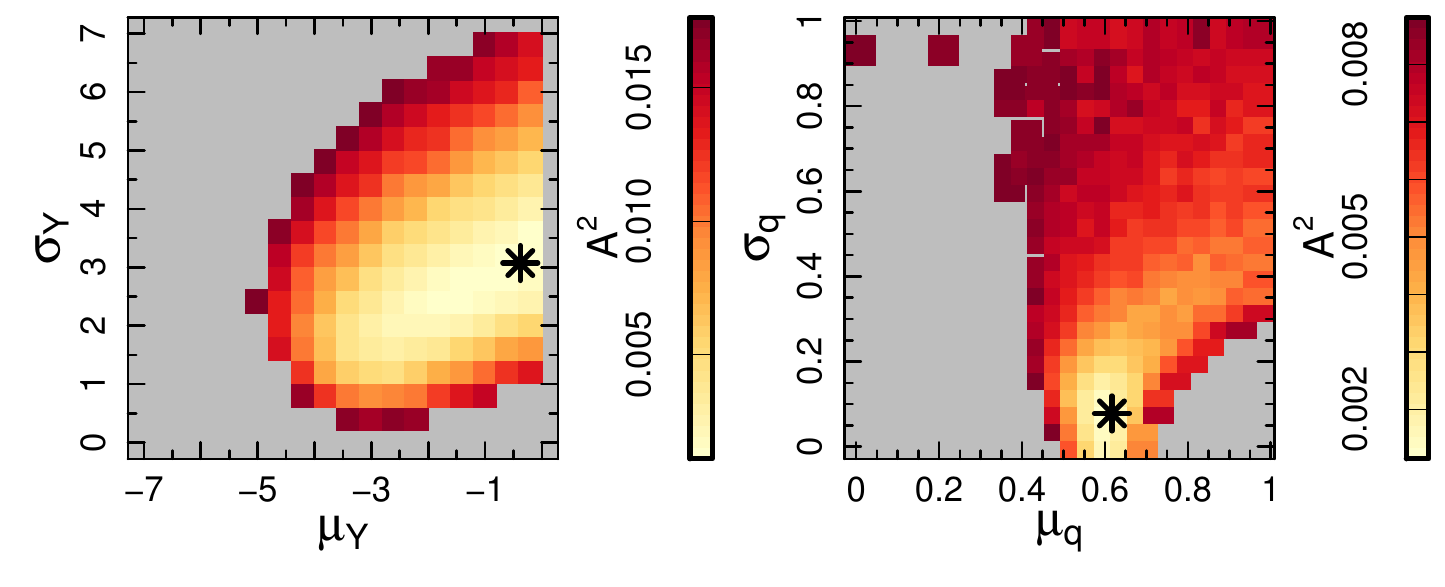}
\caption{The intrinsic shape of fast (left) and slow (right) rotators. The observed distribution of kinematic misalignments $\Psi$ and global apparent ellipticity $\epsilon$ is shown in the top panels with the observed (blue) and fitted (orange) smoothed and normalised distributions $F(\Psi)$ and $F(\epsilon)$. In the central panels, distributions shown in orange with thick and thin lines represent the 68 and 95\%\ probability intervals, respectively. The lower panels show the $A^2$ values in $Y$ vs $\sigma_Y$ space computed by fixing $q$ and $\sigma_q$ to their best fit value or $q$-$\sigma_q$ space (keeping $Y$ and $\sigma_Y$ fixed at their fitted value). In the lower panels, the best fit values are shown as a black asterisk and corresponding colour scales are shown. The distributions of data and fit lines are markedly different for the two samples, confirming that fast and slow rotators must have different intrinsic shape distributions.}\label{fig:shape_FRSR}
\end{center}
\end{figure*}

\begin{figure*}
\begin{center}
\includegraphics[width=80mm]{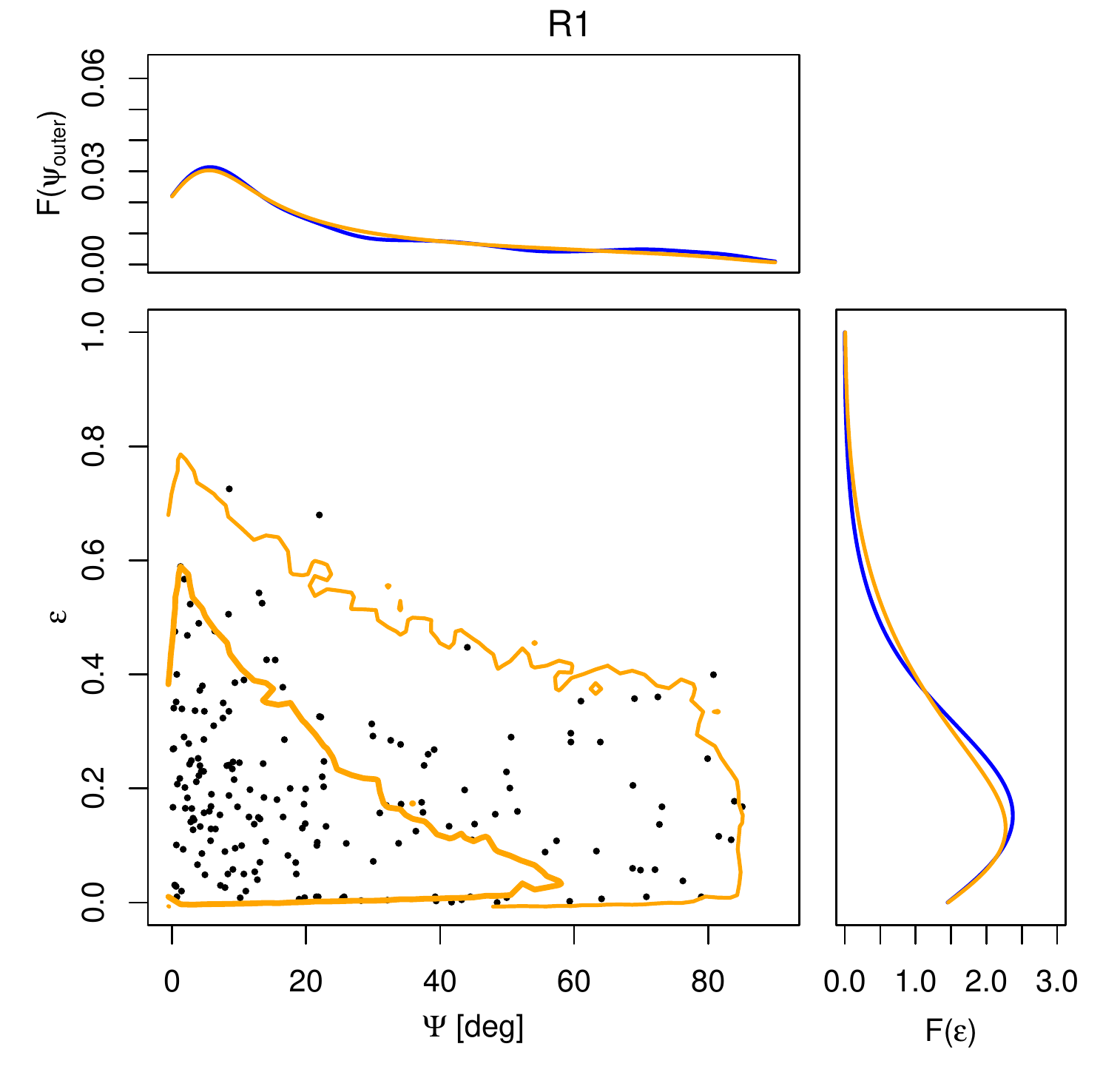}\includegraphics[width=80mm]{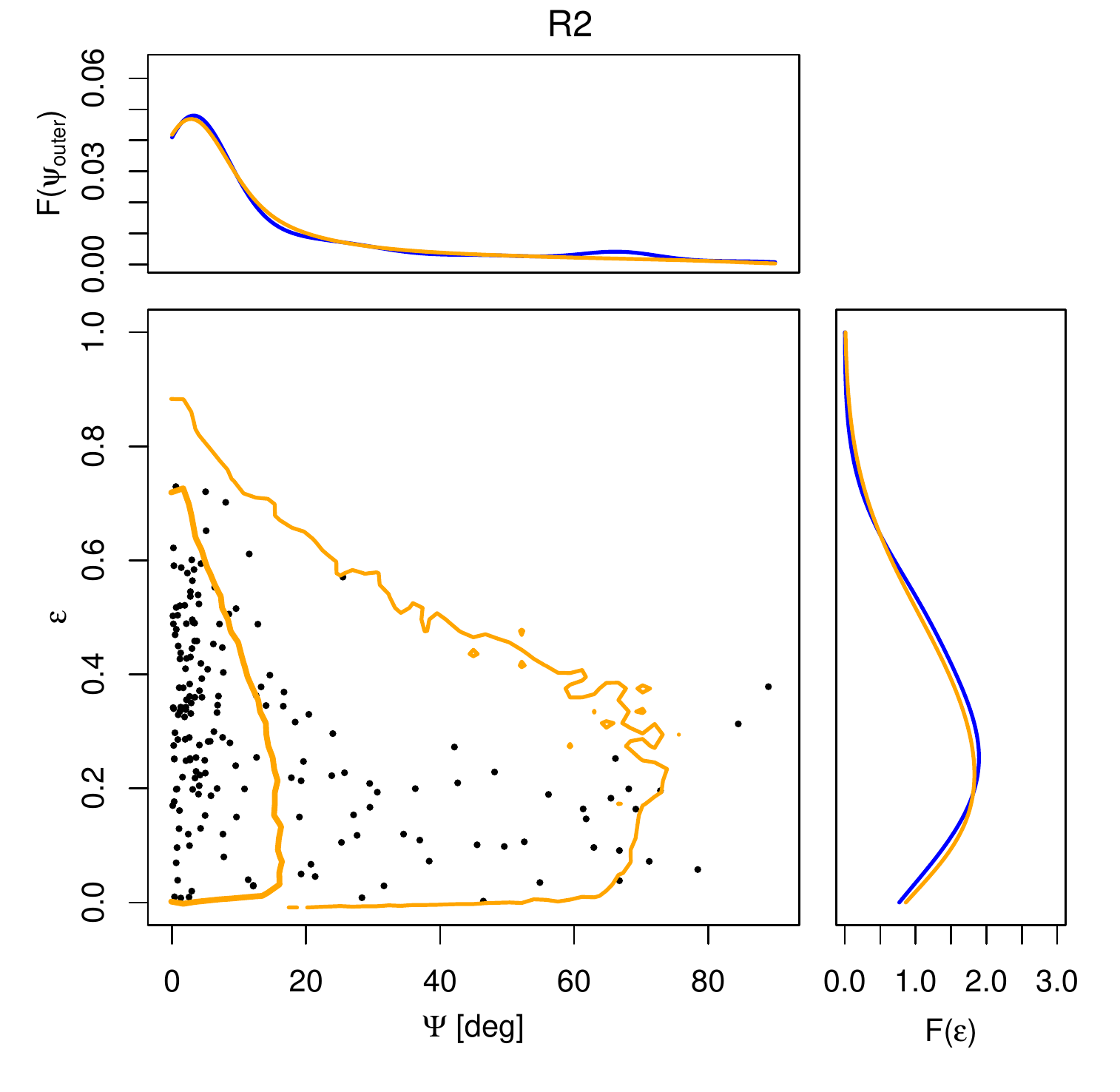}\\
\includegraphics[width=80mm]{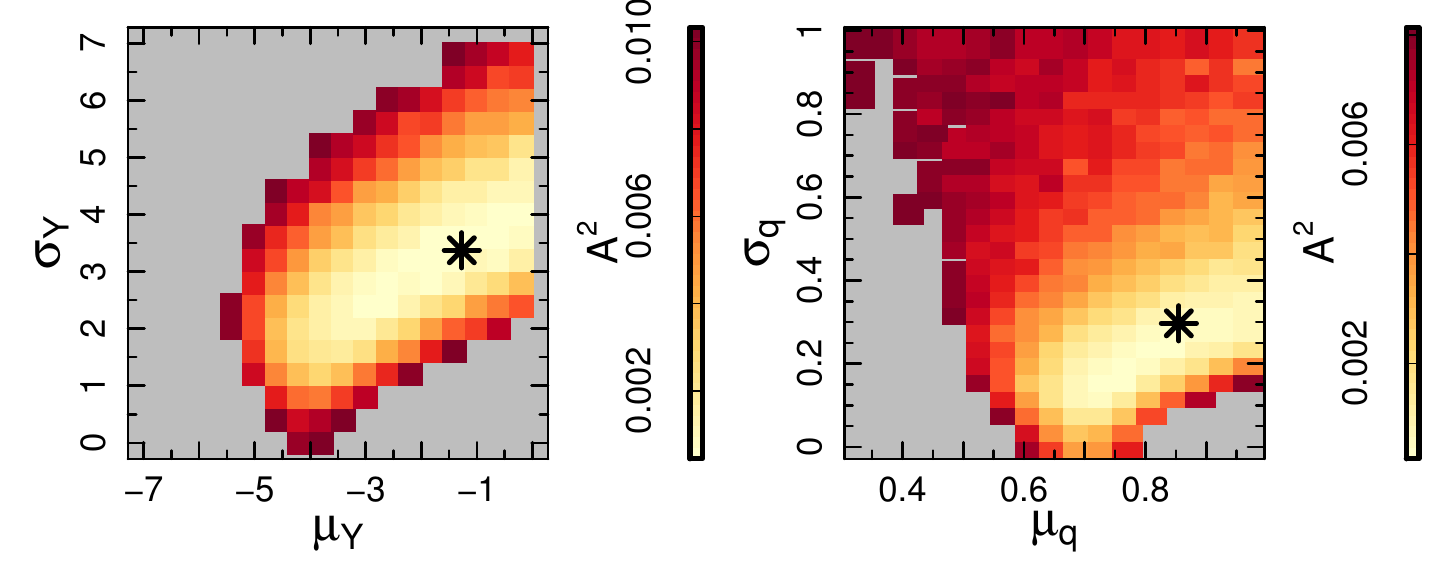}\includegraphics[width=80mm]{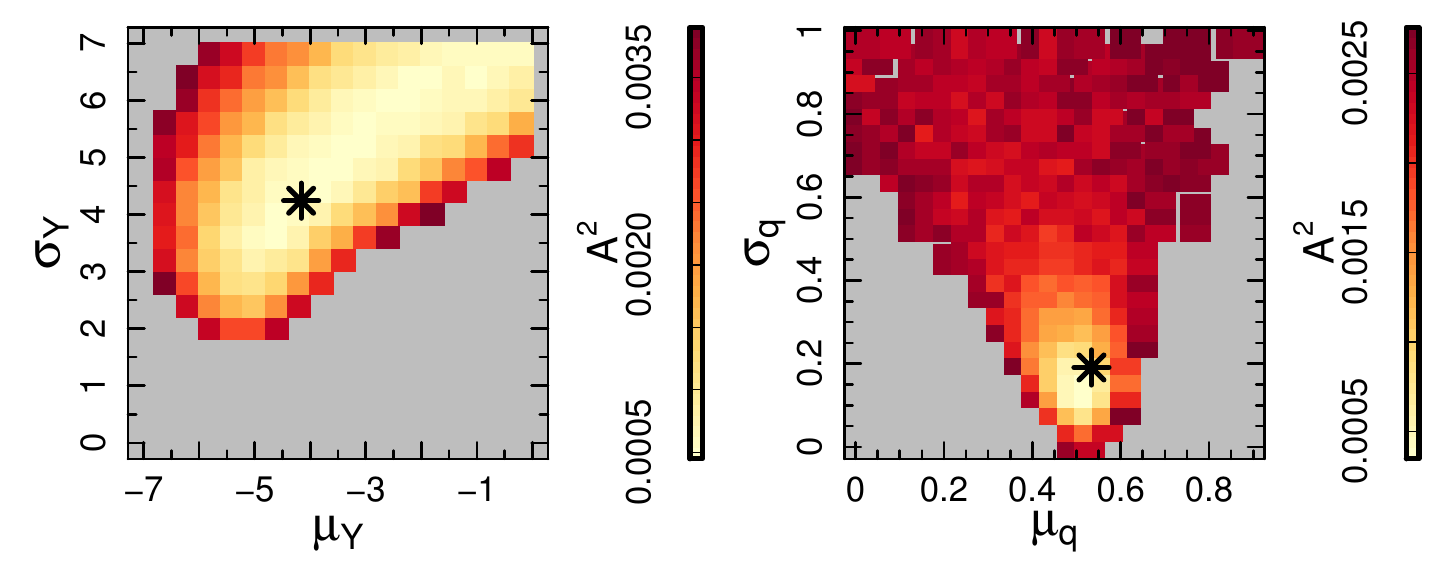}\\
\includegraphics[width=80mm]{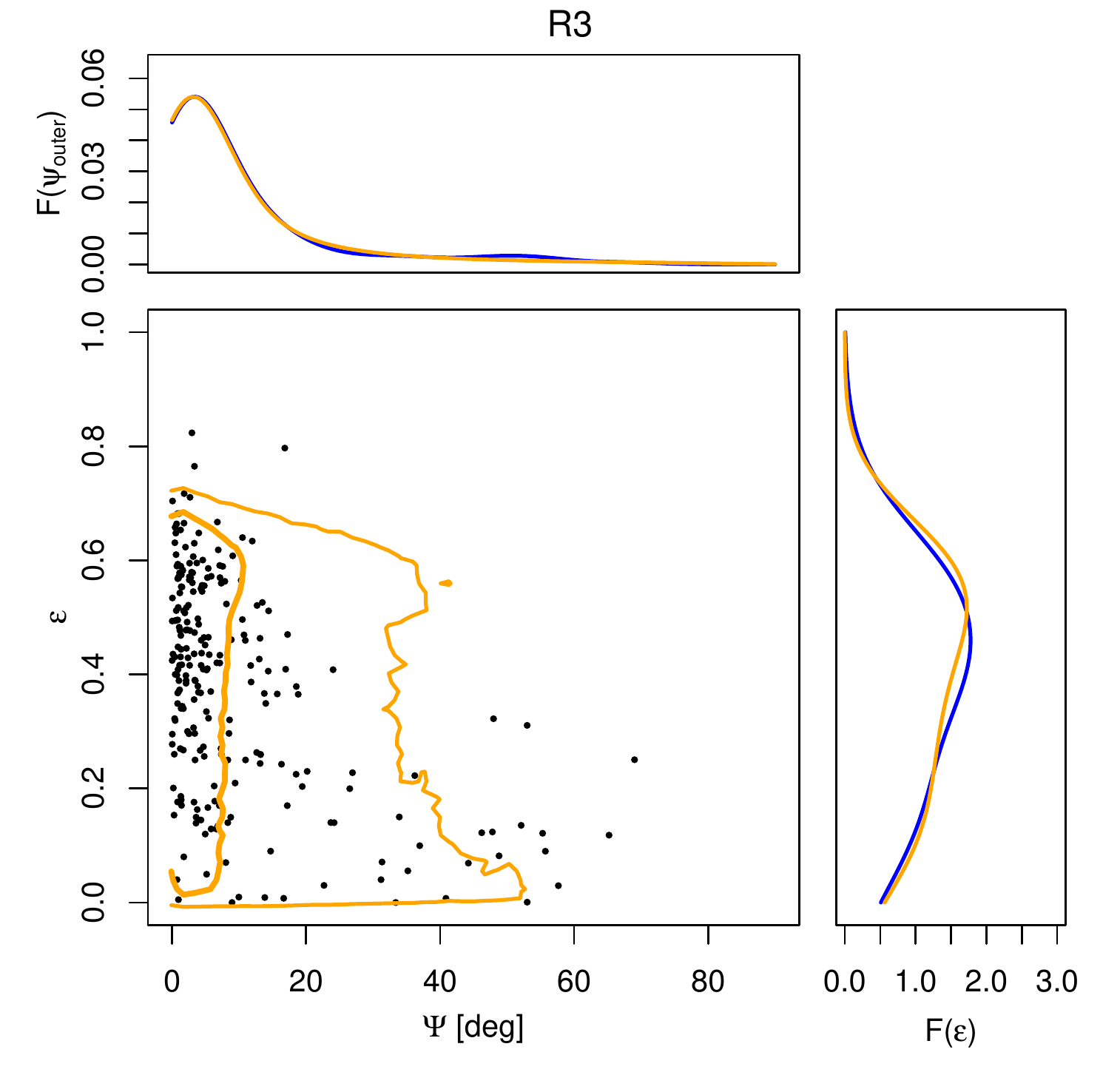}\includegraphics[width=80mm]{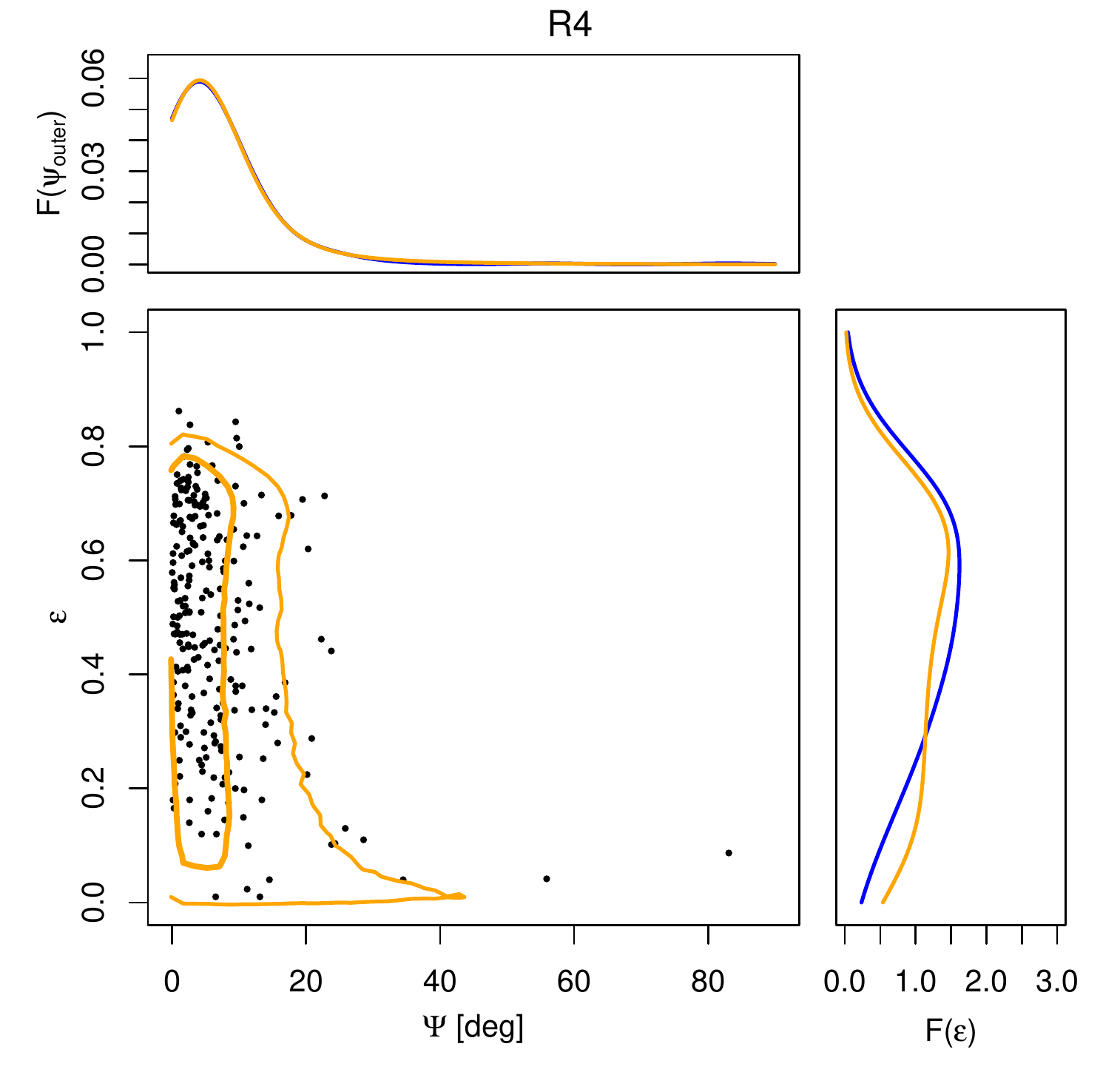}\\
\includegraphics[width=80mm]{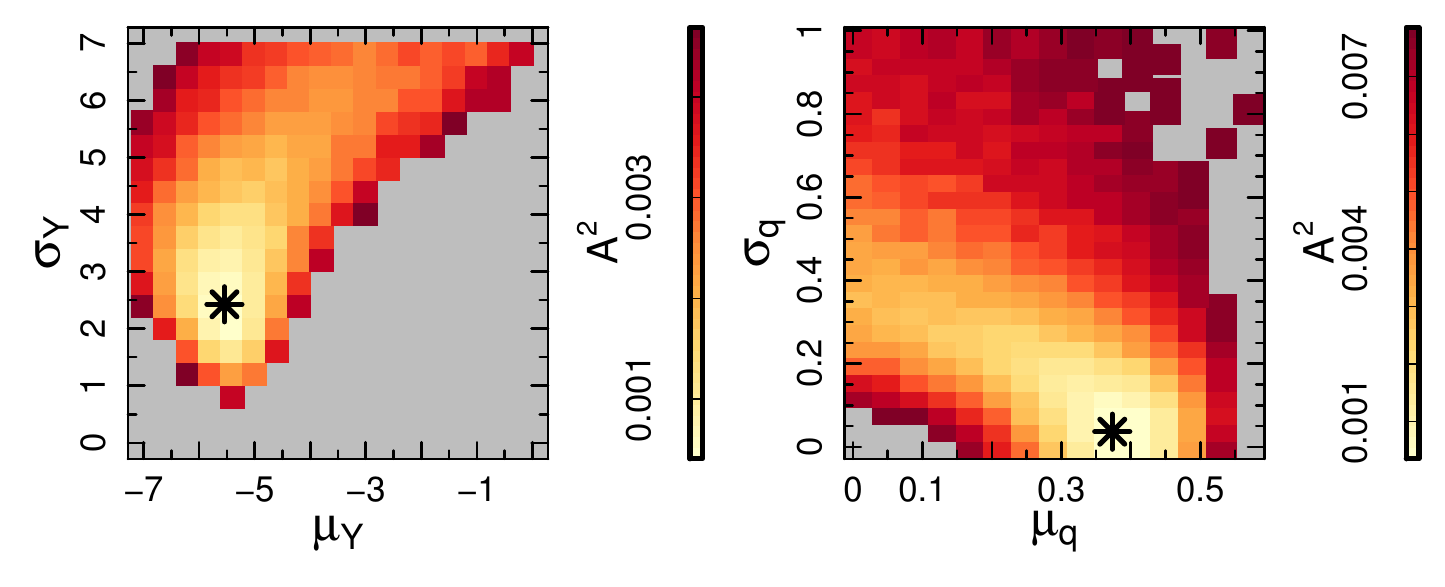}\includegraphics[width=80mm]{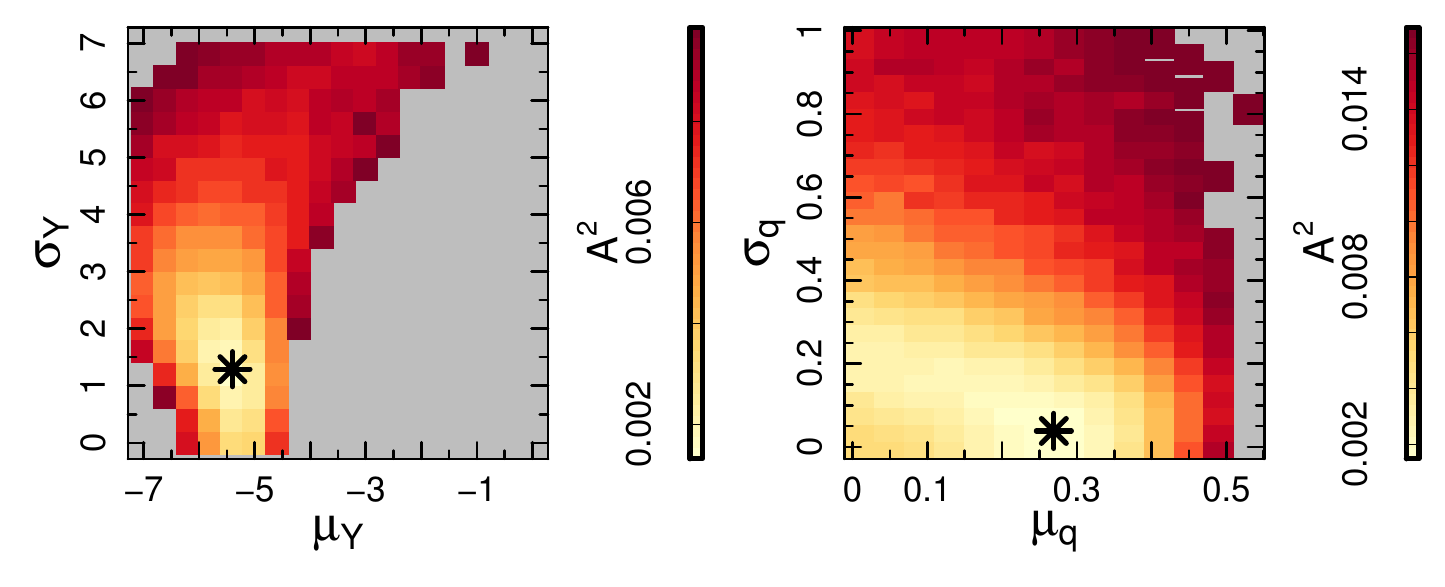}\\
\caption{The intrinsic shape of fast rotator subsamples: R1 to R4, as labelled. Panels, symbols, lines and colours as per Fig. \ref{fig:shape_FRSR}.}\label{fig:shape_subFRSR}
\end{center}
\end{figure*}

We assume that all galaxies in the sample are drawn from the same distributions of intrinsic axis ratios. The $p$ axis ratio is assumed to be log-normal $Y=\ln{(1-p)}$ with mean $\mu_Y$ and standard deviation $\sigma_Y$ following \citet{Padilla08}. A normal distribution is assumed for $q$ with mean $\mu_q$ and standard deviation $\sigma_q$. In this framework, every observed system is a random projection of a three-dimensional ellipsoid drawn from these intrinsic distributions. We marginalise over the viewing angles to eliminate the parameters $(\varphi,\nu)$. The probability of observing projection angle values $(\varphi,\nu)$ is the ratio of the area element over the total area of the unit sphere of viewing angles:
\begin{equation}
P(\varphi,\nu)=\frac{\sin\nu}{4\pi}.
\end{equation}

We follow \citet{Weijmans14} and assume that $\Psi_{\rm int}$ coincides with the viewing direction that generates a round apparent ellipticity. This is mathematically equivalent to:
\begin{equation}\label{eq:thetaint}
\tan(\Psi_{\rm int})=\sqrt{\frac{T}{1-T}}.
\end{equation}
In other words, the intrinsic misalignment depends solely on the shape of the galaxy, with values of $\Psi_{\rm int}$ largest for triaxial systems \citep[see][their appendix A, for justification]{Weijmans14}.

To fit for the intrinsic shape, we follow the method of \citet{Foster16}. Modelled distributions ($F_{\rm mod}$) are compared with the observed distributions ($F_{\rm obs}$) for $\Psi$ and $\epsilon$. The area under $F_{\rm mod}$ and $F_{\rm obs}$ is normalised to unity. In order to find the set of ($\mu_Y,\sigma_Y,\mu_q,\sigma_q$) that best matches observations, we minimise the square of the area between the modelled and observed distributions:

\begin{dmath}\label{eq:chisq} 
A^2 =  \sum_{i} (F_{\rm obs}(\Psi_{i})-F_{\rm mod}(\Psi_{i}))^2 (\delta \Psi_{i})^{2} + \\ \sum_{j} (F_{\rm obs}(\epsilon_{j})-F_{\rm mod}(\epsilon_{j}))^2 (\delta \epsilon_{j})^{2}.
\end{dmath}

Equation \ref{eq:chisq} is different from the equation used in \citet{Foster16} for optimisation, but it has the advantage of being more robust and corresponds to a more intuitively meaningful quantity. The global minimum for Equation \ref{eq:chisq} is efficiently found using the {\sc DEoptim} R package using the principles of differential evolution \citep[see][for more detail]{Mullen11}. We set a maximum of 200 iterations for each sub-sample. To minimise running time, we choose a threshold at $A^2<0.0005$, which corresponds to a total area of $A<0.02$ between the modelled and observed distributions. This threshold was chosen based on visual inspection of the fits and the typical minimum $A^2$ values reached after 200 iterations. The threshold value is somewhat arbitrary and should vary between studies depending on the data, sample size and desired precision. Variables may range within the following bounds: $-7\le\mu_Y\le0$, $0\le\sigma_Y\le7$ and $0\le\mu_q\le1$, $0\le\sigma_q\le1$. This generous range ensures convergence away from the boundary.

Before discussing the results of the shape analysis on our various samples, we need to point out possible caveats and limitations of our data, as we also outline our solutions in what follows. 
\begin{enumerate}
\item Because SAMI galaxies are on average further away than previous samples (e.g. ATLAS$^{\rm 3D}$), this affects the accuracy with which $PA_{\rm phot}$ and $\epsilon$ can be measured. Additionally, the comparatively larger seeing ($\sim 2$ arcsec) smooths small localised kinematic fluctuations. For these reasons, the uncertainties on the observed parameters is expected to be larger than in \citet{Weijmans14}. In order to mitigate the effects of measurement uncertainties on our discrete distributions, we smooth the observed and modelled distributions of points using Gaussians with standard deviations $\sigma_{\Psi}=5^{\circ}$ and $\sigma_{\epsilon}=0.1$ for the kinematic misalignment and apparent ellipticity distributions, respectively. 
\item A large proportion of lenticular and spiral galaxies have a bar \citep[e.g.][]{Marinova07,Barway11}. Bars can skew our measurements of $PA_{\rm phot}$ and $\epsilon$. Our photometry is optimised to minimise the effects of bars. A visual inspection reveals a small range in fraction of objects with photometry clearly affected by a bar (3\% and 5\%) for all samples. We thus conclude that contamination from bars should be small.
\item We verify that selecting galaxies with stellar kinematics has not biased our distribution of apparent ellipticities. The lower left panel of Fig. \ref{fig:sample_properties} shows that the distributions in ellipticities are similar between the whole SAMI and the parent sample with the notable difference of an excess of galaxies at $\epsilon\sim0$ in the former. This excess is caused by a large number of galaxies that are small on the sky and hence because the ellipticities cannot be measured they default to zero (as discussed above in relation to MGE fitting). We run a KS-test on both ellipticity distributions for $\epsilon>0.05$ yields a p-value of 0.9999 indicating that both distributions are consistent with having been drawn from the same parent distribution at a very high significance level. Furthermore, the SAMI target selection did not consider apparent ellipticity and should thus be free of such biases.
\item Finally, \citet{Padilla08} show that dust extinction affects the distribution of measured apparent ellipticities in magnitude selected samples because edge-on (i.e. high ellipticity) dusty galaxies are more likely to drop out of the sample than face-on galaxies as they suffer higher extinction. \citet{Padilla08} found that the effect of dust was measurable at a ``low statistical significance'' in their magnitude selected sample of over 500,000 galaxies. Our sample is several orders of magnitudes smaller than the \citet{Padilla08} study, so we cannot meaningfully correct for such a small effect. More importantly, the correction suggested by \citet{Padilla08} does not apply here because the SAMI target selection is based on highly complete stellar mass cuts \citep[see][their figure 4]{Bryant15} rather than magnitudes. Based on equation 3 of \citet{Bryant15}, we do not expect our stellar mass estimates to be strongly affected by dust/inclination because the colour and magnitude terms nearly cancel each other out.
\end{enumerate}

\section{Analysis and results}\label{sec:analysis}

\subsection{Measured intrinsic shapes}

Our algorithm is first run on the sample of fast rotators. This is the largest sample studied in this work and therefore the least likely to be plagued by stochastic effects. The observed and fitted distributions of kinematic misalignments and apparent ellipticities are shown in Fig. \ref{fig:shape_FRSR}. Fitted values are listed in Table \ref{table:shapesoutput}. After 4 iterations, the fit converges to a near perfect fit with only $A^2=0.0004$. The $A^2$ maps in Fig. \ref{fig:shape_FRSR} and Appendix \ref{sec:allA2maps} show a clear minimum in the 4 dimensional parameter space. The modes of the axis ratio distributions indicate that fast rotators are typically axisymmetric oblate spheroids with intrinsic flattening $\mu_q=0.32$.

\begin{figure}
\begin{center}
\includegraphics[width=80mm]{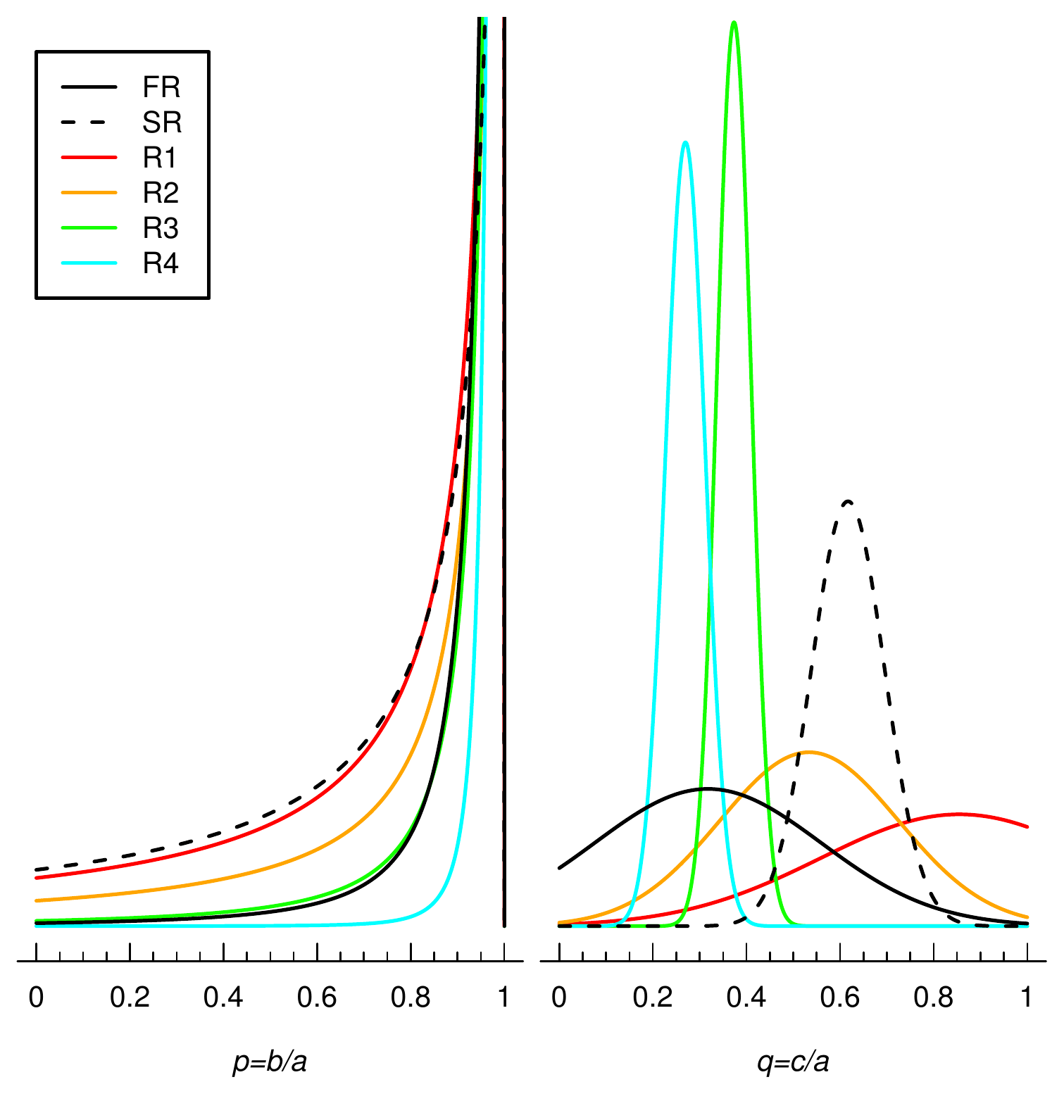}
\caption{Distributions of fitted intrinsic axis ratios $p$ and $q$ for the slow rotators (dashed line), fast rotators (solid black line), R1(red), R2 (orange), R3 (green) and R4 (cyan). All distributions peak around $p\sim1$ indicating that most galaxies in SAMI are axisymmetric. The ``spread'' of $p$ values is distinctly broader for R1 galaxies than for R 2-4 galaxies, indicating a higher fraction of triaxial (i.e. $p\ne1$) galaxies in the former. Similarly, slow rotators have a higher fraction of triaxial galaxies than fast rotators.}\label{fig:RatiosDistr}
\end{center}
\end{figure}

The procedure is repeated for the slow rotators and the other kinematic samples. Results are shown in Figs. \ref{fig:shape_FRSR} and \ref{fig:shape_subFRSR}. Fig. \ref{fig:RatiosDistr} summarises the various fitted axis ratio distributions for all samples. The samples of fast rotators and R1-3 converge. The slow rotator and R4 samples fail to converge although the global minimum in $A^2$ is successfully identified (see $A^2$ maps in Fig. \ref{fig:shape_subFRSR} and Appendix \ref{sec:allA2maps}). Possible reasons for non-convergence are discussed below. Typical galaxies in \emph{all samples} are found to be majoritively oblate spheroids ($p\sim1$) with varying intrinsic flattening. For slow rotators and R1 galaxies, higher values of $\mu_Y$ combined with large $\sigma_Y$ indicate a larger proportion of triaxial systems when compared to fast rotators. 
There is a clear and monotonous trend of decreasing fraction of triaxial systems between R1 and R4. While these trends are robust, we do not quote exact fractions because those are uncertain and ultimately depend on the assumed shape of the axis ratio distributions (normal vs lognormal). In all samples, the majority of galaxies are axisymmetric with the intrinsic flattening varying in inverse proportion to their rotational support.


\begin{figure*}
\begin{center}
\includegraphics[width=150mm]{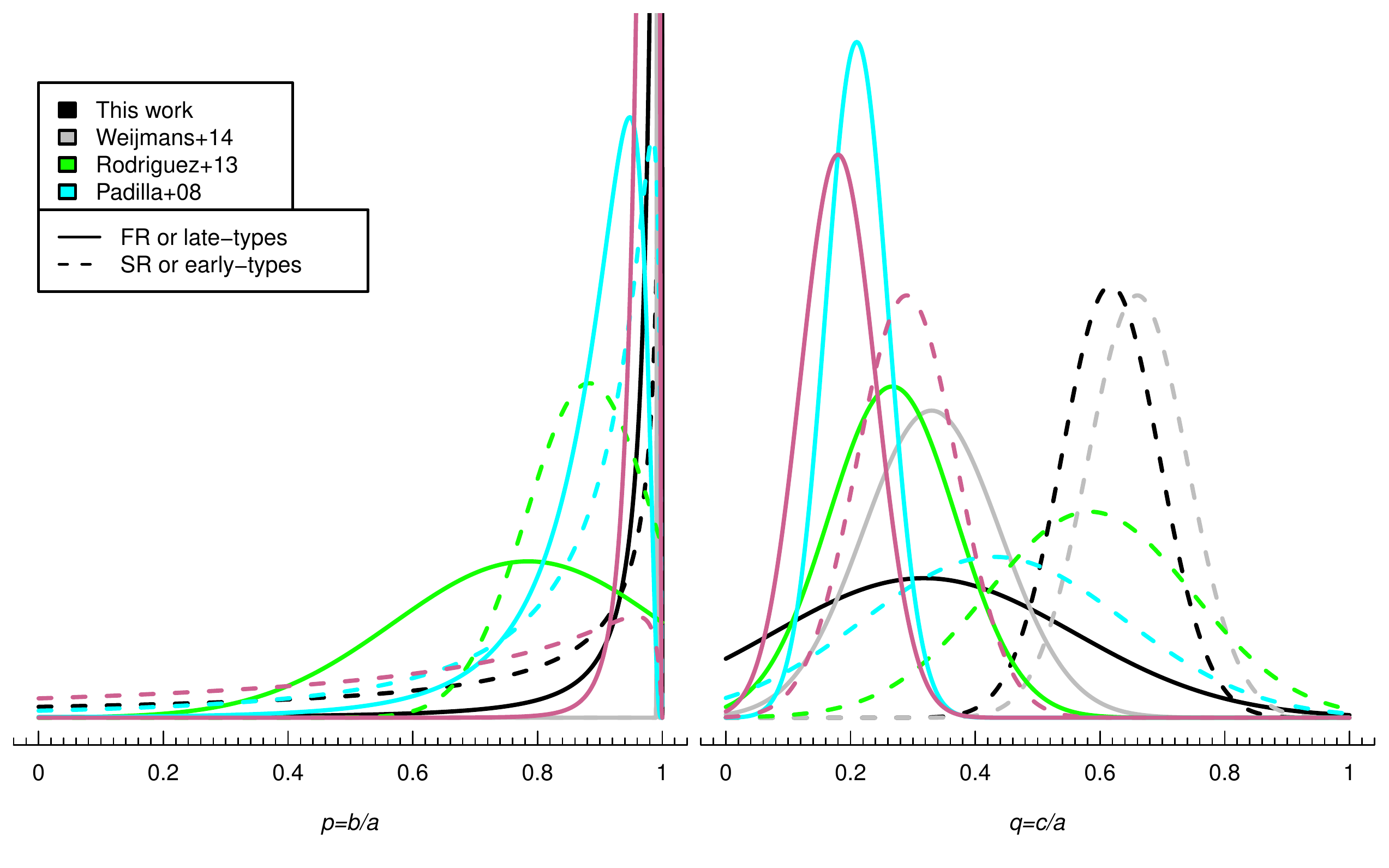}
\caption{Comparison of the axis ratio distributions for slow (dashed lines) and fast (solid lines) rotators in the ATLAS$^{\rm 3D}$ \citep[][grey]{Weijmans14} and this work (black). Although presented here for completeness, the apparent ellipticity and kinematic misalignment distributions could not be reliably reproduced in \citet{Weijmans14} for slow rotator galaxies. While the distribution of intrinsic flattening ($q$) overlap and generally agree, the axis ratio $p$ distributions look very different. In particular, the ATLAS$^{\rm 3D}$ results do not allow for any significantly triaxial ($p\ne1$) galaxies (see text and appendix). \emph{While not directly comparable due to distinct methods and sample selection,} other selected literature results \citep{Rodriguez13, Padilla08,Ryden06} based on inverting the distribution of apparent ellipticity alone for early (dashed coloured lines) and late (solid coloured lines) type samples are also shown as labeled.}\label{fig:litcomp}
\end{center}
\end{figure*}

Overall, the relative shapes of the various samples are consistent with what is expected given their respective rotational support, even though not all samples reach convergence. We identify possible reasons for the lack of convergence for slow rotators and R4 galaxies in order of perceived likelihood. 
\begin{enumerate}
\item The sample of SR is our smallest sample and hence the measured $\Psi$ and $\epsilon$ distributions are noisier, leading to a higher $A^2$.
\item Sample R4 contains relatively few round galaxies in projection because rotation appears lower for face-on views. This may translate into an orientation bias.
\item Based on previous work \citep[e.g.][]{Krajnovic11} and on the distribution of S\'ersic indices in Fig. \ref{fig:sample_properties}, the SR sample is likely a inhomogeneous blend of truly pressure supported systems and flattened axisymmetric non-regular rotators, such as galaxies hosting a central kinematically decoupled core or 2$\sigma$ kinematic feature unresolved by SAMI.
\item Our assumed $\Psi_{\rm int}$ (Equation \ref{eq:thetaint}) may not be applicable for R4. If the model is inappropriate, the data cannot be explained by the model, which leads to a larger $A^2$ value. This is discussed further in what follows.
\end{enumerate}

Because galaxies in R4 are rotationally supported, an alternative assumption is that the intrinsic angular momentum vector is aligned with the short axis $\Psi_{\rm int}=0$. In other words, the apparent kinematic misalignment is only large for projections near face-on. Running the algorithm with this latter assumption yields slightly higher $A^2=0.004$ values with a lower fraction of triaxial systems (lower $\sigma_Y$) and nearly identical $q$ distributions. 

As with \citet{Weijmans14} and due to computational limitations, we are unable to provide uncertainties on our measured intrinsic shape parameters. Moreover, uncertainties are not available for our photometric measurements and hence measurement errors cannot be propagated directly. The smoothing we have applied to both the $\Psi$ and $\epsilon$ distributions however is quite large (i.e. $\sigma_{\Psi}=5^{\circ}$ and $\sigma_{\epsilon}=0.1$) compared to the expected observational uncertainties. This indicates that the ``noise'' in the distributions is dominated by stochastic uncertainties rather than by measurement uncertainties. Since the same smoothing is applied to both the model and the data self-consistently, it is safe to assume that photometric uncertainties do not significantly contribute to the shape parameters uncertainties. To get a handle on the order of magnitude of the uncertainties on the various shape parameters, we refer to the $A^2$ maps in Fig. \ref{fig:shape_subFRSR} and \ref{fig:shape_FRSR}, which show a range of fitted parameters with comparably good fits. The $A^2$ maps suggest that for the fast rotators, R3 and R4 samples, values of $\mu_Y$ and $\sigma_Y$ within $\sim1-2$ of the best fit parameters provide similarly good fit to the data, while $\mu_q$ and $\sigma_q$ are more uncertain due to degeneracies towards lower $\mu_q$ and higher $\sigma_q$. For slow rotators, R1 and R2, the reverse is true, the $A^2$ maps suggest that $\mu_Y$ and $\sigma_Y$ are degenerate towards lower $\mu_Y$ and $\sigma_Y$, while $\mu_q$ and $\sigma_q$ within $\sim0.1-0.2$ of the best fit parameters provide similarly good fit to the data.

In summary, we infer meaningful intrinsic shapes for the majority of our samples (Table \ref{table:shapesoutput}). SAMI galaxies are typically oblate axisymmetric spheroids with slow rotators and R1 samples containing a significant fraction ($\sim15\%$) of triaxial galaxies. We reliably measure that fast rotators are intrinsically more flattened than slow rotators. We show that, as spin (i.e. $\lambda_{R_e}$) increases, galaxies become intrinsically flatter and less likely to be triaxial. 

\subsection{Comparison with literature}

Before discussing the implications of our findings for the formation and evolution of galaxies in the different samples, we compare our intrinsic shape findings with that previously measured in the literature. The work of the ATLAS$^{\rm 3D}$ team \citep{Weijmans14} on the topic of intrinsic shapes offers the only comparable study to date as it also used sizeable samples of galaxies with spatially resolved kinematic maps to measure the kinematic position angle and a comparable method \citep{Franx91}. A notable difference between this work and that of \citet{Weijmans14} is the absence of late-type (spiral) galaxies in the latter. The distributions of intrinsic axis ratios for fast and slow rotators measured by \citet{Weijmans14} are compared to those found in this work in Fig. \ref{fig:litcomp}. We note that our fit for slow rotators has $A^2=0.0006$, slightly higher than our threshold value. 

In their work, \citet{Weijmans14} found an intrinsic flattening distribution of $\mu_q=0.33$, $\sigma_q=0.08$ with $\mu_Y=-5$ and $\sigma_Y=0.08$ for their sample of fast rotators. The fitted value of $\mu_Y$ for the ATLAS$^{\rm 3D}$ fast rotators corresponds the boundary of the grid searched. The authors also comment that the value of $\sigma_Y$ was largely unconstrained. A $\sigma_Y$-value of 0.08 is very small, corresponding to an extremely narrow range in allowed $p$ values. Even for rotationally supported galaxies, it is unlikely that they nearly all have perfectly circularised discs as suggested by these results. \citet{Weijmans14} discuss some tension between their results and previous literature results for spiral and early-type galaxies as the fast rotators are much closer to perfect axisymmetry. The results presented here for the SAMI fast rotators also indicate a lower degree of axisymmetry than reported by \citet{Weijmans14}; however, a more generous range of $p$ values is preferred owing to the more generous parameter bounds explored. We emphasise that our sample is not morphologically selected, and hence differs fundamentally from that of ATLAS$^{\rm 3D}$, which contains early-type galaxies only. As such, it is not surprising that the intrinsic axis ratios do not match exactly between the two samples. In Appendix \ref{appendix:ATLAS3D} we perform our intrinsic shape analysis on the ATLAS$^{\rm 3D}$ data. Our results indicate a broader range in intrinsic flattening for the SAMI fast rotators with a more skewed distribution of disc circularity than reported in \citet{Weijmans14} due to differences in the implementation of the method.

The intrinsic shape of galaxies has also been constrained for large samples by inverting the distribution of apparent ellipticities only. These studies are more common as they only require photometric information. Due to the lack of stellar kinematics, the samples in these studies are not directly comparable to those used here, which were kinematically selected. We emphasise that while late-type galaxies do rotate faster on average than early-type galaxies, fast and slow rotators are not equivalent to ellipticals and lenticulars/spirals \citep[see][for an extensive discussion]{Emsellem07}. Hence, the following (non-exhaustive) comparison, while arguably instructive, is limited; and we include it mainly for completeness. 

\citet{Ryden06} performed the shape analysis on late-type and early-type galaxies in the 2MASS galaxy survey. Based on $K$-band images, the intrinsic axis ratio distributions for the late-type sample were $\mu_Y=-3.86$, $\sigma_Y=0.74$, $\mu_q=0.18$ and $\sigma_q=0.06$. For the early-type sample, axis ratios were $\mu_Y=-0.14$, $\sigma_Y=1.74$, $\mu_q=0.29$ and $\sigma_q=0.08$. 

The SDSS presents a wealth of galaxy images for this analysis to be performed on. \citet{Padilla08} infer intrinsic axis ratio distributions of $\mu_Y=-2.3$, $\sigma_Y=0.79$, $\mu_q=0.21$ and $\sigma_q=0.05$ for late-type spirals and $\mu_Y=-2.2$, $\sigma_Y=1.4$, $\mu_q=0.21$ and $\sigma_q=0.05$ for their early-type sample. 

The results from photometric studies are summarised in Fig. \ref{fig:litcomp}. For both \citet{Ryden06} and \citet{Padilla08}, the intrinsic flattening is usually lower for their early-type (late-type) sample than for our slow (fast) rotator sample. This makes sense since 1) early-type galaxies include flattened lenticular galaxies which would normally be considered fast rotators, and 2) late-type galaxies are the fastest rotators \citep{Fogarty14}. \citet{Rodriguez13} find $q$ distributions more consistent with the present work ($\mu_q=0.267$, $\sigma_q=0.102$ for spiral galaxies and $\mu_q=0.584$, $\sigma_q=0.164$ for ellipticals), but they fit Gaussian distributions to $p$ instead of lognormal distributions, so our $p$ distributions cannot be compared directly. 

Most previous studies find that the mode of the $p$ axis ratio distribution is close to axisymmetry (i.e. $p_{\rm mode}\sim1$) with varying degrees of skewness to lower values. While the various literature samples are not directly comparable to the ones used here, the results of this work sit comfortably within the range found from previous studies.

\section{Discussion}\label{sec:discussion}

\subsection{The intrinsic shape of kinematically selected galaxies}

We fit the observed distributions of kinematic misalignments ($\Psi$, Equation \ref{eq:psi}) and apparent ellipticities ($\epsilon$) for the fast and slow rotators and four kinematic sub-samples in the SAMI Galaxy Survey using the method of \citet{Franx91}. Our results (summarised in Table \ref{table:shapesoutput}) compare favourably with previous galaxy intrinsic shape measurements from the literature. 

\begin{figure}
\begin{center}
\includegraphics[width=80mm]{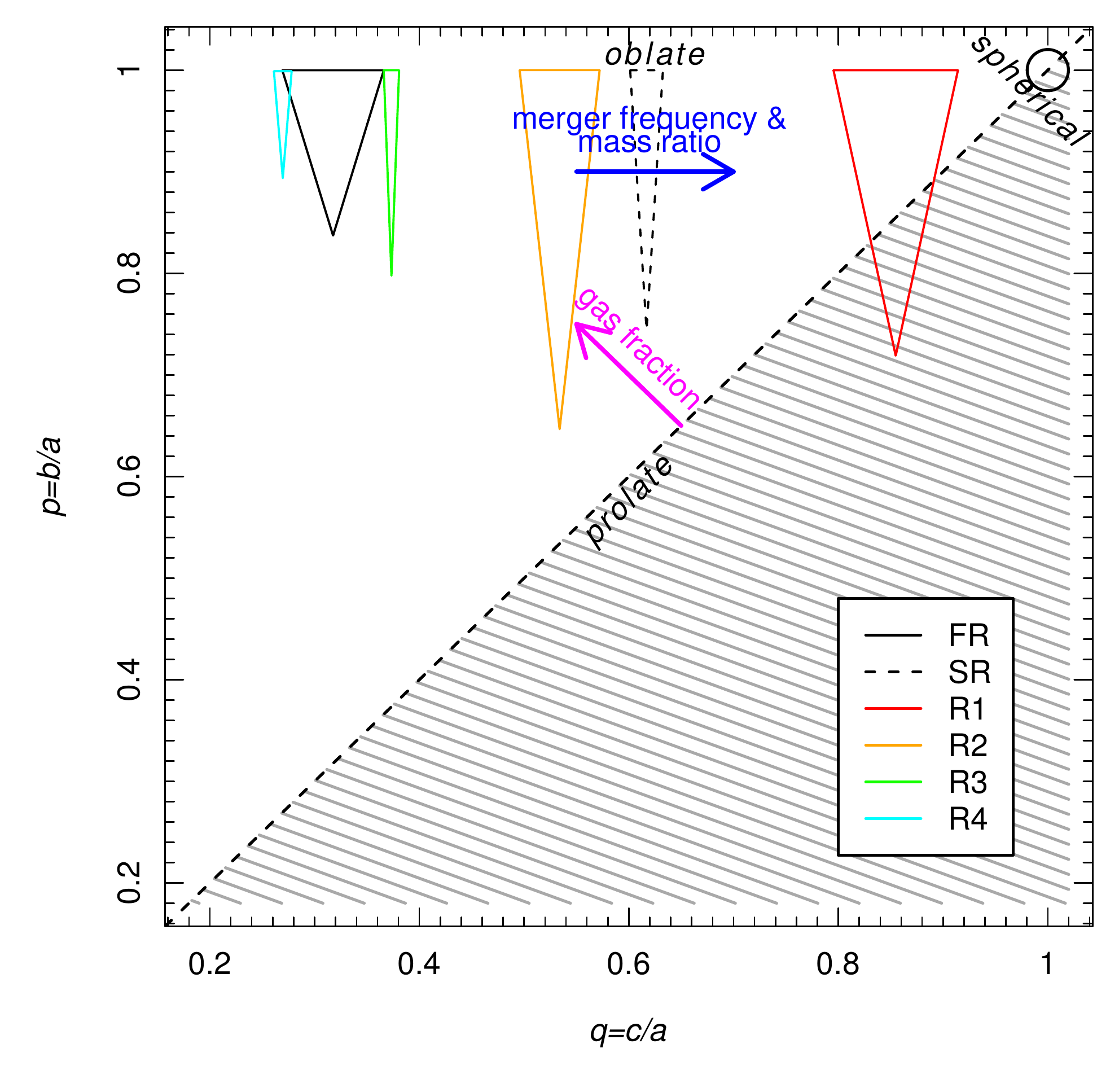}
\caption{Summary of the intrinsic axis ratios of fast rotators (solid black), slow rotators (dashed line), R1 (red), R2 (orange), R3 (green) and R4 (cyan) galaxies. For each sample, the centre of the base of the triangle is located at the mode of the distribution with the width and length proportional to $\sigma_q$ and $sigma_Y$, respectively. The positions of perfectly oblate ($p=1$), prolate ($p=q$) and spherical ($p=q=1$) ellipsoids are shown. Galaxies in the SAMI Galaxy Survey are typically very closely axisymmetric oblate with varying degrees of intrinsic flattening depending on their rotational support. The magenta arrow shows the expected effect of increasing the gas fractions in binary mergers \citep{Cox06}. The blue arrow shows the effect of increased merger frequency and merger mass ratio on the intrinsic shape based on \citet{Jesseit09} and \citet{Moody14}.}\label{fig:TheoreticalSummary}
\end{center}
\end{figure}

We find that typical galaxies in the SAMI Galaxy Survey are oblate axisymmetric with varying degrees of intrinsic flattening ($q=c/a$). In agreement with our slow rotator sample, R1 galaxies have inferred intrinsic $p=b/a$ distributions that are skewed to lower values (see Fig. \ref{fig:RatiosDistr}) indicating that these samples contain a non-negligible fraction of triaxial galaxies ($p\ne1$). R1 galaxies have the highest $q$ values of all the samples studied in this work. 

R2 galaxies have some rotational support and are thus slightly more flattened and axisymmetric than R1 galaxies. Their intrinsic flattening is lower than that of slow rotators in SAMI and they are on average more axisymmetric.

Convergence could not be reached for R4 galaxies, leaving a relatively poorer fit to the intrinsic shape for this sample, although the global minimum was successfully identified. Both R3 and R4 galaxies have high spin parameter. They have very similar intrinsic shapes and are on average more axisymmetric and more flattened than R1, R2 and slow rotator galaxies.

Fast rotators show a wide range of intrinsic flattening values, although on average fast rotators are more flattened than slow rotators. The $p$ axis ratio distribution for fast rotators in the SAMI Galaxy Survey is intermediate between that of R2 and R3 galaxies.

\subsection{The effect of rotation on intrinsic shapes}

In Fig. \ref{fig:FRSR_selection}, we show the theoretical position of an edge-on axisymmetric galaxy with anisotropy parameter $\beta=0.7\epsilon_{\rm intr}$ as a magenta line. The majority of galaxies ($>85$\%) in our samples are either axisymmetric or very nearly so ($p>0.8$). It is thus reasonable to expect that the typical galaxy in each sample (i.e. median $\lambda_{R_e}$ and intrinsic ellipticity $\epsilon_{\rm intr}=1-q$) should lie on this magenta line. While the typical galaxy in R1 and R4 lie almost exactly on that line, R2 and R3 are offset to higher intrinsic ellipticity values. This could indicate that galaxies have slightly higher anisotropies than initially inferred (i.e. $\beta>0.7\epsilon_{\rm intr}$). However, the depth of the global minimum around $\mu_q$ shown in Fig. \ref{fig:shape_subFRSR} suggests that values within $\Delta \mu_q\sim\pm0.05$ of that would still yield a reasonable $A^2$ value, indicating that R2 and R3 essentially agree with the magenta line.

As expected from the magenta line in Fig. \ref{fig:FRSR_selection}, we confirm that the measured intrinsic flattening inversely correlates with rotational support such that galaxies that rotate faster tend to be more flattened. The $p$ distributions of R1 to R4 galaxies show decreasing levels of skewness, indicating systematically increasing fractions of axisymmetric systems as rotational support increases.

\subsection{Comparison with simulations}

We now compare our results to theoretical expectations for the intrinsic shapes of galaxies. Unfortunately, while most theoretical work on the topic gives qualitative descriptions of the intrinsic shapes of e.g. merger remnants, the quantitative values are usually not given. For this reason, we can usually compare our intrinsic shape results to simulations in qualitative terms. The mass ratio of galaxy merger progenitors plays a role in shaping the remnant galaxy. \citet{Jesseit09} and \citet{Moody14} found that minor mergers led to flatter remnants (lower $q$) and higher triaxiality than major mergers. The most triaxial galaxies are usually formed by sequential mergers or re-mergers \citep{Moody14}. Similarly, \citet{Taranu13} found that multiple dry minor merger remnants usually led to triaxial systems. Published histograms of intrinsic ellipticities indicate that the intrinsic axis ratios varied between $0.5\lesssim p\lesssim1$ and $0.5\lesssim q\lesssim 0.8$ with skewed and symmetric distributions, respectively. Using binary major mergers simulations, \citet{Cox06} found that dissipational (or gas rich) mergers typically led to more oblate and flattened remnants. In other words, galaxies that form with large degrees of dissipation tend to have higher $p\sim1$ and lower $q$ values. The effects of these formation scenarios are summarised in Fig. \ref{fig:TheoreticalSummary} where the absolute position of the various scenarios is approximated based on the qualitative descriptions from \citet{Cox06}, \citet{Jesseit09} and \citet{Moody14}.

In this framework, the intrinsic shape of fast rotators and R3-4 galaxies is consistent with low mass ratio merger histories and dissipative star formation (i.e. gas rich mergers). By contrast, slow rotators and R1 galaxies have intrinsic shapes consistent with dissipationless (i.e. gas-poor) major mergers or multiple dissipationless minor mergers. R2 galaxies are consistent with intermediate scenarios.

Linking our results to the shape of the underlying dark matter potential is non-trivial given that many factors are likely involved in shaping galaxies. The simulations of \citet{Bailin07} showed that disc galaxies embedded in triaxial dark matter haloes would have slightly elliptical discs with the most elliptical discs corresponding to the most triaxial haloes. Considering that the fastest rotating systems (R3 and R4 galaxies) are dominated by spiral and lenticular galaxies and ignoring the effect of bars on the ellipticity of the disc, the distribution of $p$ axis ratio may reflect the triaxiality of the dark matter haloes in which spiral galaxies reside. The distribution of intrinsic $p$ axis ratios for the R3 and R4 galaxies allows for a considerable fraction of galaxies with slightly elliptical discs ($p<0.95$ or $\epsilon_{\rm intr}>0.05$) of order up to $\sim10$ percent suggesting that triaxial dark matter haloes may be common.

\section{Conclusions}\label{sec:conclusion}

We successfully invert the distributions of apparent ellipticities and kinematic misalignments to infer the intrinsic shape of kinematically selected galaxy samples in the SAMI Galaxy Survey following the method of \citet{Franx91}.

We empirically demonstrate that
\begin{enumerate}
\item most galaxies in the SAMI Galaxy Survey are oblate axisymmetric.
\item Fast rotators are distinctly more intrinsically flattened than slow rotators and the latter present a significant fraction ($15$\%) of triaxial systems.
\item As theoretically expected, the intrinsic shape of galaxies is a strong function of rotational support. We demonstrate that samples of increasingly high ``spin'' parameter proxy ($\lambda_{R_e}$) exhibit higher degrees of intrinsic flattening (i.e. lower $q$) and higher fractions of axisymmetric systems (i.e. $p\sim1$). To our knowledge, our work is the first to statistically and simultaneously confirm these trends observationally.
\item Comparison to galaxy formation models suggest that the intrinsic shape of high-spin galaxies is consistent with high gas fractions, low merger frequencies and mass ratios. Conversely, low-spin galaxies have intrinsic shapes consistent with multiple dissipationless mergers.
\end{enumerate}

The SAMI Galaxy Survey is ongoing and set to double in sample size by the time of its completion in 2018. This work is the first in a series to explore the inter-dependence of galaxy intrinsic shapes and other intrinsic properties.

\section*{Acknowledgments}
We thank R. Davies for insightful discussions.
The SAMI Galaxy Survey is based on observations made at the Anglo-Australian Telescope. The Sydney-AAO Multi-object Integral field spectrograph (SAMI) was developed jointly by the University of Sydney and the Australian Astronomical Observatory. The SAMI input catalogue is based on data taken from the Sloan Digital Sky Survey, the GAMA Survey and the VST ATLAS Survey. The SAMI Galaxy Survey is funded by the Australian Research Council Centre of Excellence for All-sky Astrophysics (CAASTRO), through project number CE110001020, and other participating institutions. The SAMI Galaxy Survey website is http://sami-survey.org/.
JvdS is funded under Bland-Hawthorn's ARC Laureate Fellowship (FL140100278). SMC acknowledges the support of an Australian Research Council Future Fellowship (FT100100457). M.S.O. acknowledges the funding support from the Australian Research Council through a Future Fellowship (FT140100255). 
NS acknowledges support of a University of Sydney Postdoctoral Research Fellowship 
Support for AMM is provided by NASA through Hubble Fellowship grant \#HST-HF2-51377 awarded by the Space Telescope Science Institute, which is operated by the Association of Universities for Research in Astronomy, Inc., for NASA, under contract NAS5-26555.  
RMcD is the recipient of an Australian Research Council Future Fellowship (project number FT150100333).
SB acknowledges the funding support from the Australian Research Council through a Future Fellowship (FT140101166).

\appendix

\section{Comparison with ATLAS$^{\rm 3D}$}\label{appendix:ATLAS3D}

\begin{figure*}
\begin{center}
\includegraphics[width=80mm]{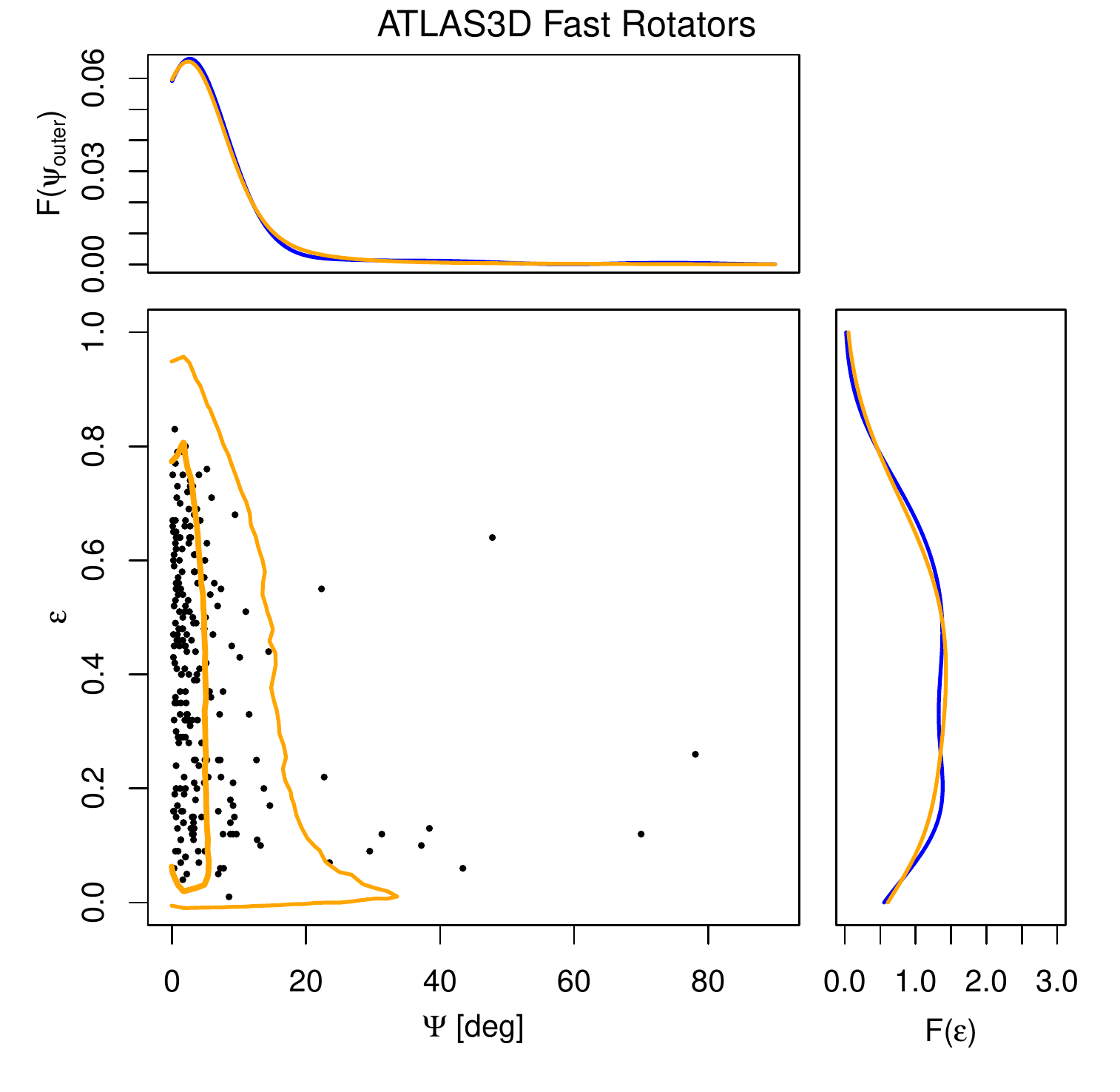}\includegraphics[width=80mm]{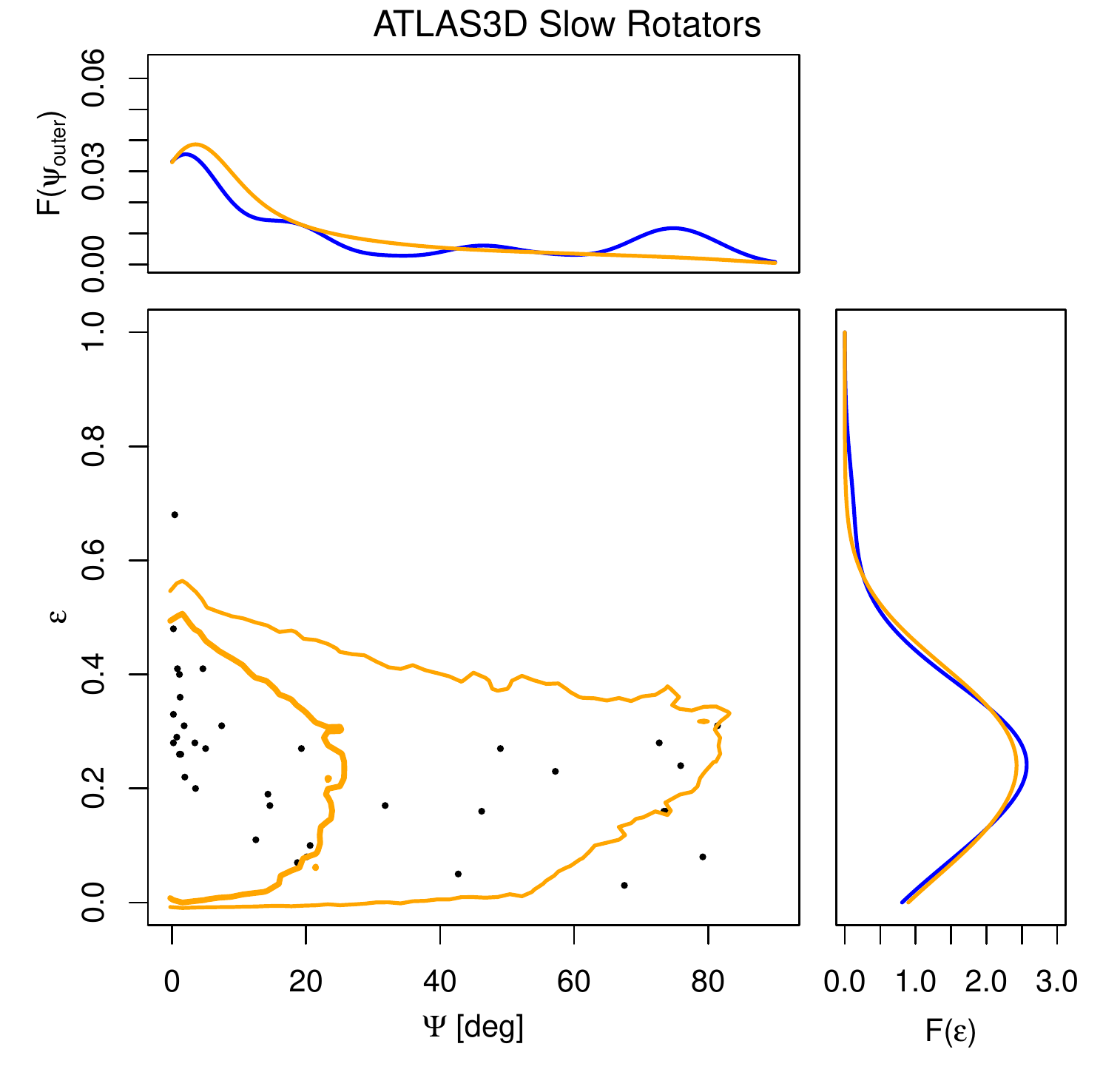}\\
\includegraphics[width=80mm]{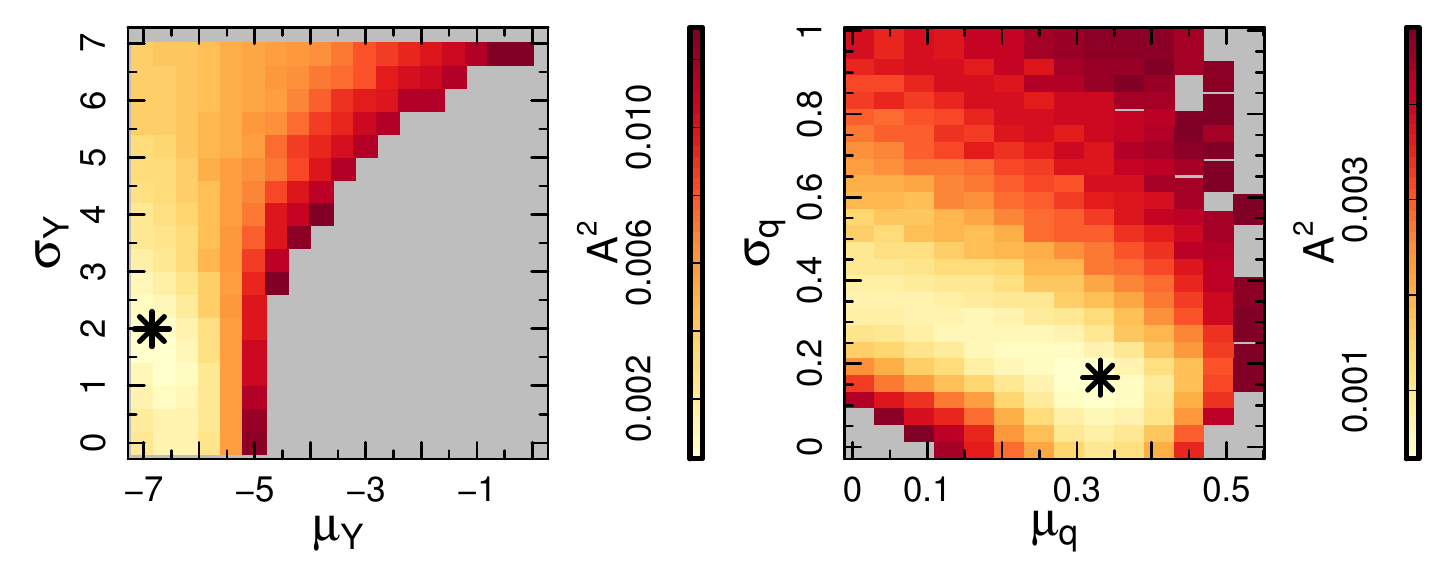}\includegraphics[width=80mm]{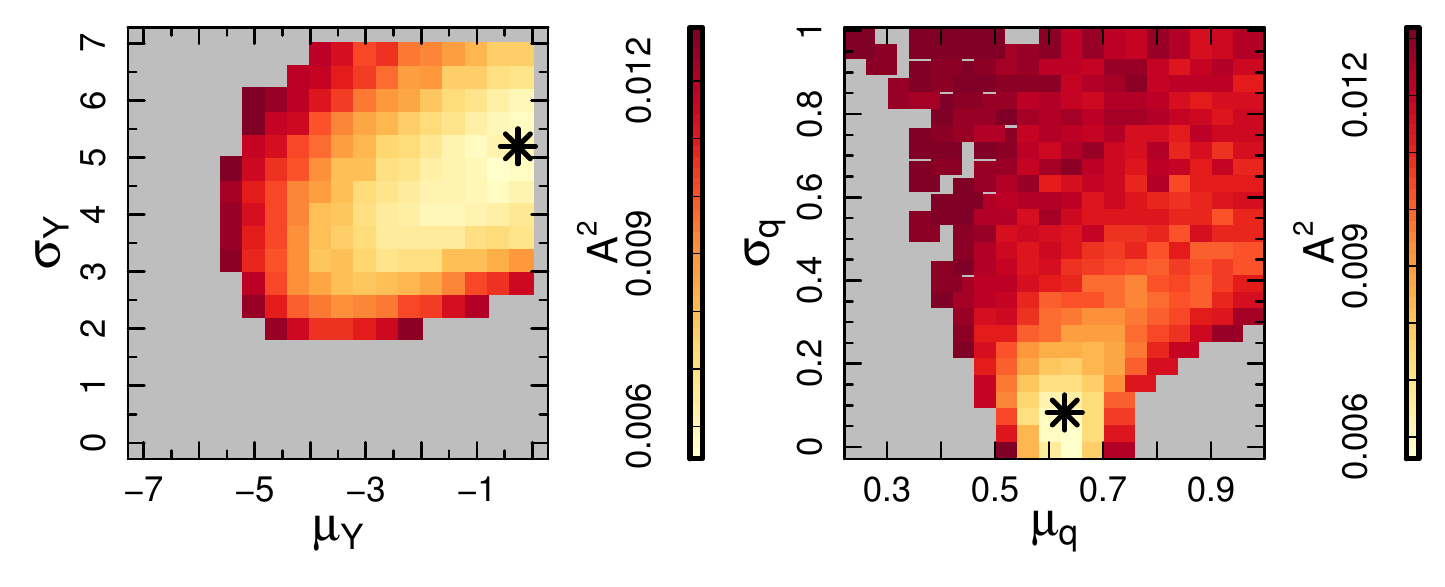}
\caption{The intrinsic shape of ATLAS$^{\rm 3D}$ fast (left) and slow (right) rotators. Panels, symbols, lines and colours as per Fig. \ref{fig:shape_FRSR}.}\label{fig:A3DFRSR}
\end{center}
\end{figure*}

\begin{figure}
\begin{center}
\includegraphics[width=80mm]{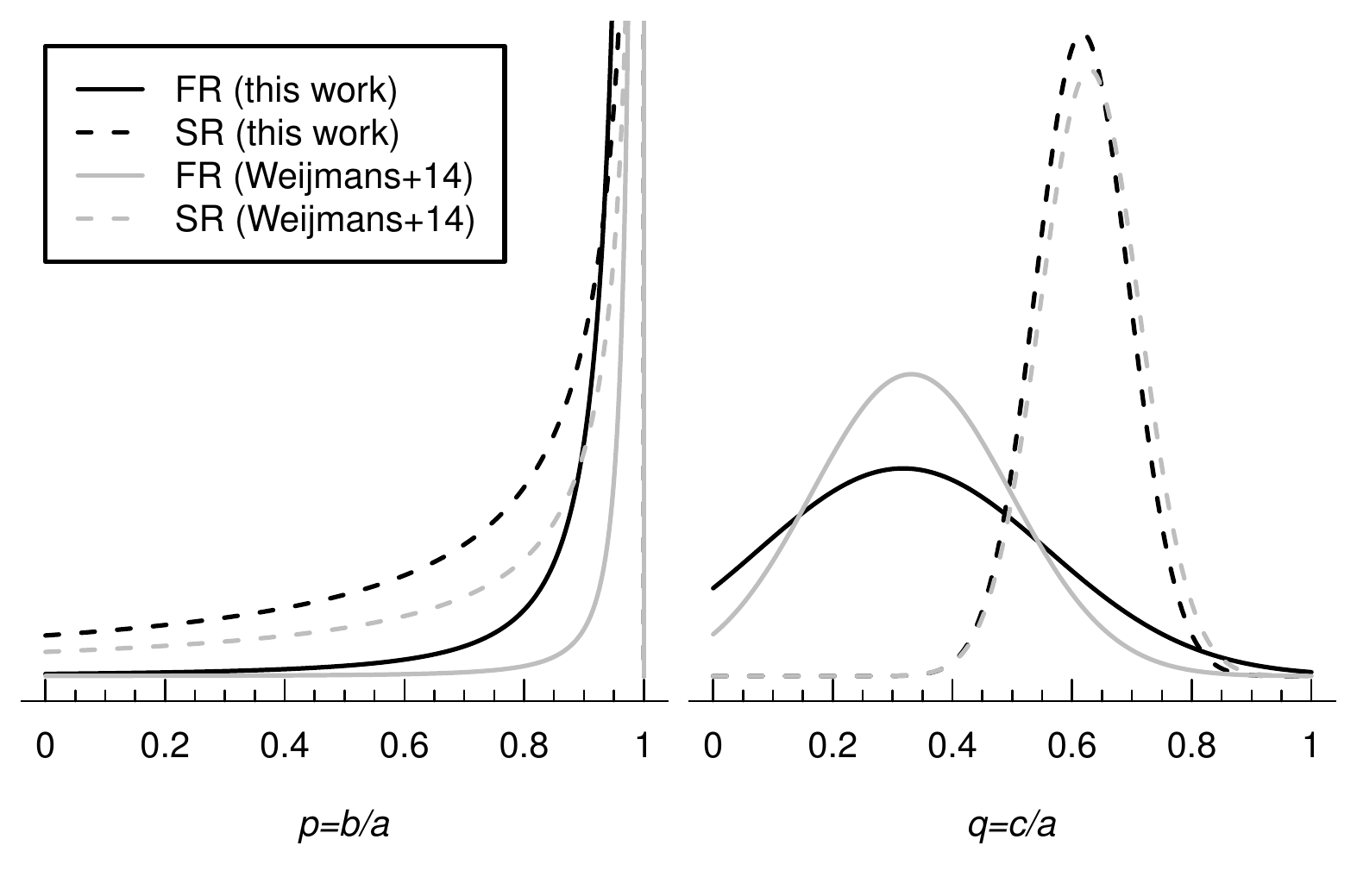}
\caption{Comparison of the axis ratio distributions for slow (dashed lines) and fast (solid lines) rotators in ATLAS$^{\rm 3D}$ (grey) and SAMI (black) as fitted with our algorithm. The agreement in the distributions of intrinsic $q$-ratio between the two samples is remarkable for both fast and slow rotators. The distributions of intrinsic disc circularity (i.e. the $p$-axis ratios) for slow rotators agree well, however fast rotators are more axisymmetric $p\sim1$ in the ATLAS$^{\rm 3D}$ sample than in SAMI.}\label{fig:A3Dcomp}
\end{center}
\end{figure}

To ensure comparability of the SAMI and ATLAS$^{\rm 3D}$ results, we also fit the distributions of kinematic misalignments and ellipticities published in \citet{Krajnovic11} with the slow ($N_{\rm gals}=36$) and fast rotator ($N_{\rm gals}=224$) samples defined in \citet{Emsellem11} using our intrinsic shape algorithm. The results are shown in Fig. \ref{fig:A3DFRSR}. For the fast rotators, we obtain $A^2=0.0003$ within 9 iterations for $\mu_Y=-6.85$, $\sigma_Y=1.99$, $\mu_q=0.33$ and $\sigma_q=0.17$, while for the slow rotators we do not reach convergence within 200 iterations with a minimum $A^2=0.005$ for $\mu_Y=-0.26$, $\sigma_Y=5.19$, $\mu_q=62$ and $\sigma_q=0.08$. Despite the higher $A^2$ value, there is a clear minimum in the $A^2$ maps in Fig. \ref{fig:A3DFRSR}. It is likely that stochastic effects as a result of the small sample size are causing the poor fit. 

In Fig. \ref{fig:A3Dcomp} we compare our fitted axis ratio distributions for the SAMI and ATLAS$^{\rm 3D}$ fast and slow rotators. There is good agreement in the distributions of $p$ and $q$ for slow rotators, despite the lack of convergence. While the intrinsic flattening distributions ($q$) agree well for fast rotators between the two samples, the ATLAS$^{\rm 3D}$ fast rotators are typically closer to perfect axisymmetry than the SAMI fast rotators. Our measured $p$ distribution is however not as narrow as that inferred by \citet{Weijmans14}, likely as a result of our more generous parameter search. Indeed, the set of best fit parameters found here is not within the bounds of the grid searched in \citet{Weijmans14}. We note that \citet{Weijmans14} also found that fast rotators in ATLAS$^{\rm 3D}$ tend to be more axisymmetric than expected from previous work on spiral galaxies \citep[e.g.][]{Padilla08}, possibly due to the inherent difficulties of measuring precise photometric values in the presence of features such as spiral arms and dust. Given that the SAMI sample of fast rotators also contains spiral galaxies (i.e. no morphological selection was applied), it may explain the broader range of $p$ values seen in our sample..pdf

\section{$A^2$ maps}\label{sec:allA2maps}

For completeness, we present the full set of low resolution $A^2$ maps for each sample fit. These highlight degeneracies as diagonal $A^2$ contours and highlight uncertainties in specific parameters through the depth of the $A^2$ contours. The most significant degeneracies are usually between $\mu_Y$ and $\sigma_Y$, and/or $\mu_q$ and $\sigma_q$. All other parameter combinations are orthogonal.

\begin{figure*}
\begin{center}
\includegraphics[width=80mm]{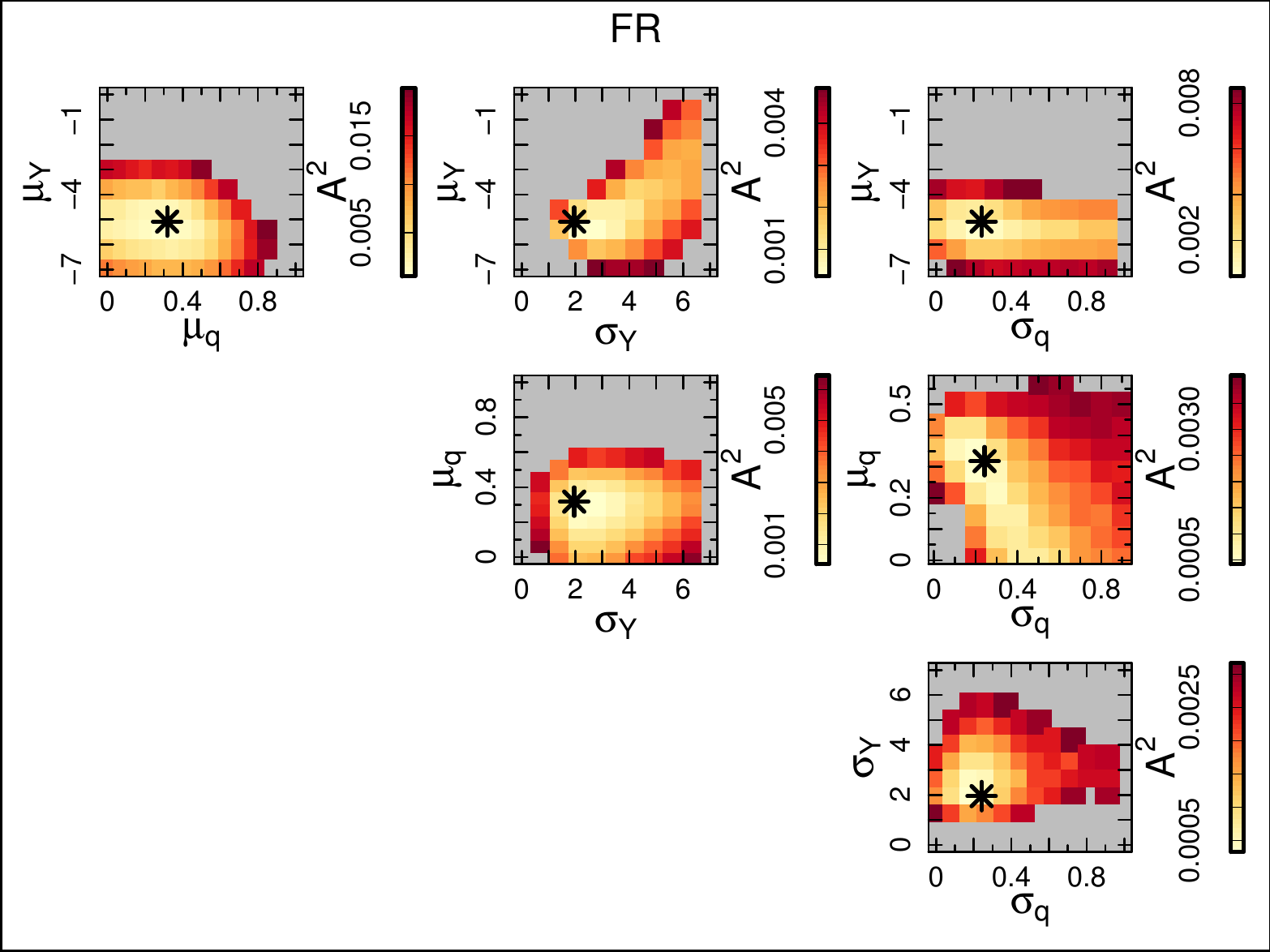}\includegraphics[width=80mm]{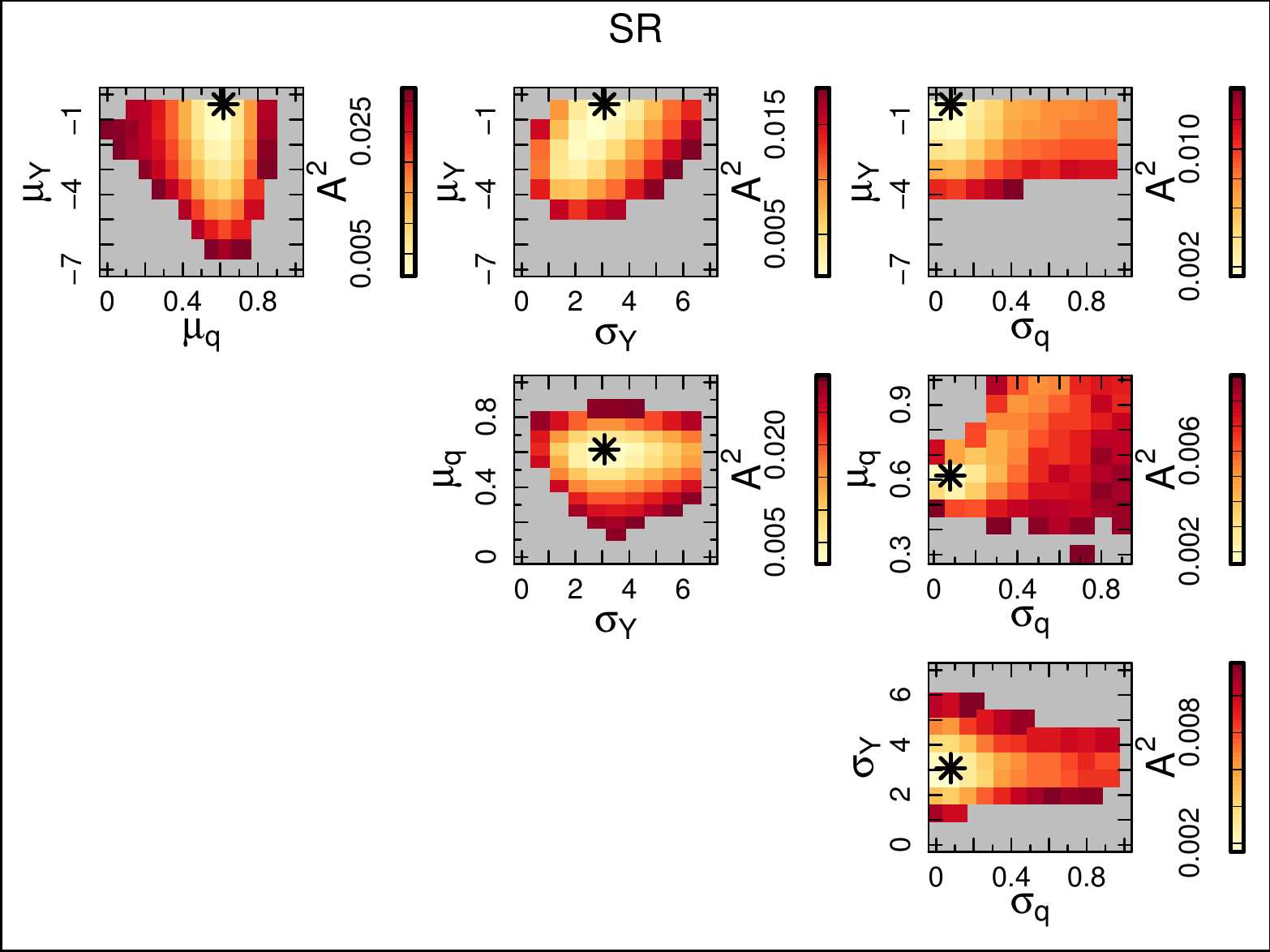}
\caption{\bf $A^2$ values in pairs of fitted parameters (e.g. $\mu_Y$ and $\mu_q$) for fast (left) and slow (right) rotators computed by fixing the other two fitted variables (e.g. $\sigma_Y$ and $\sigma_q$) to their respective best fit values. In each panel the best fit values are shown as a black asterisk and corresponding colour scales are shown.}\label{fig:allA2}
\end{center}
\end{figure*}

\begin{figure*}
\begin{center}
\includegraphics[width=80mm]{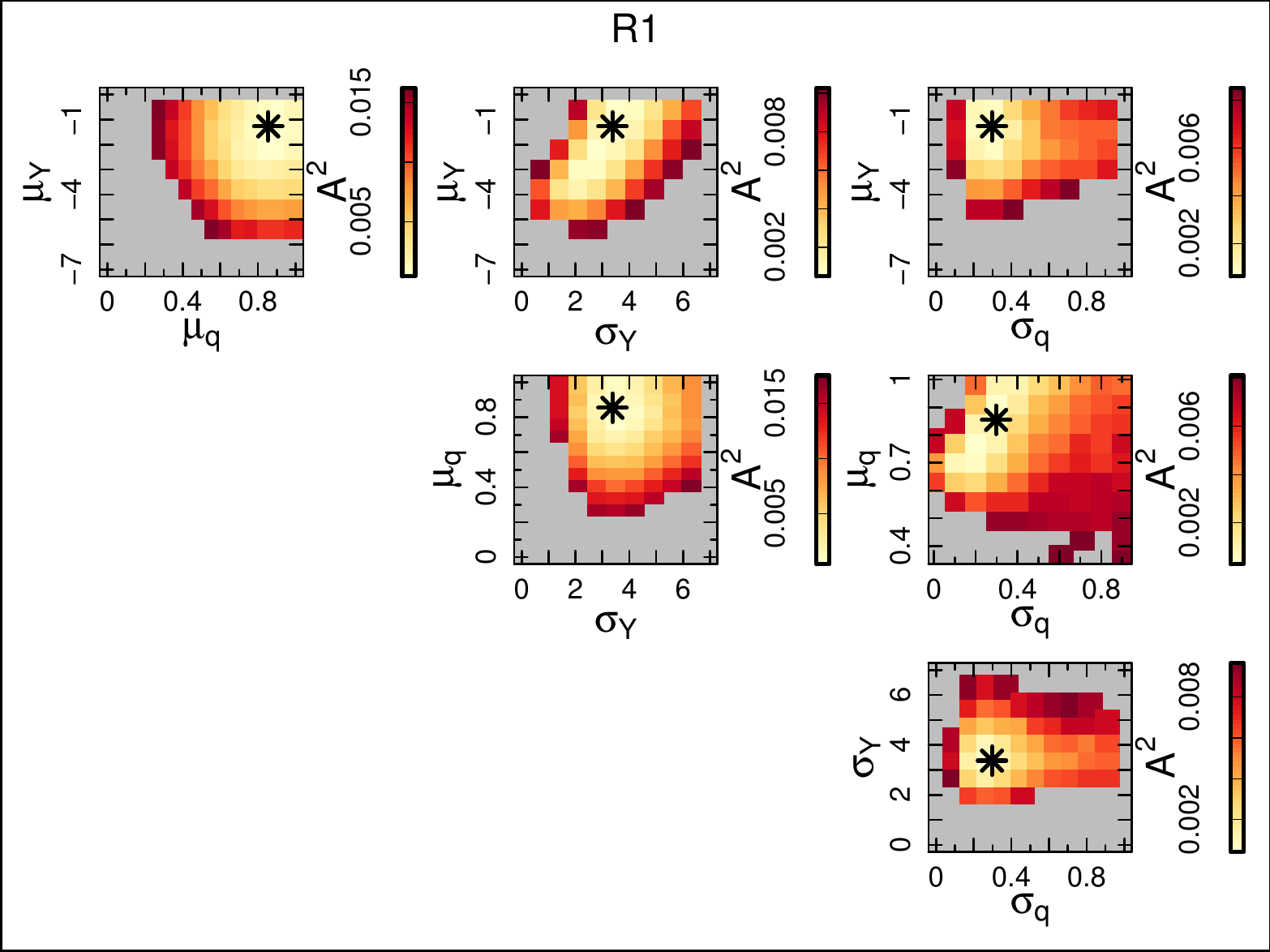}\includegraphics[width=80mm]{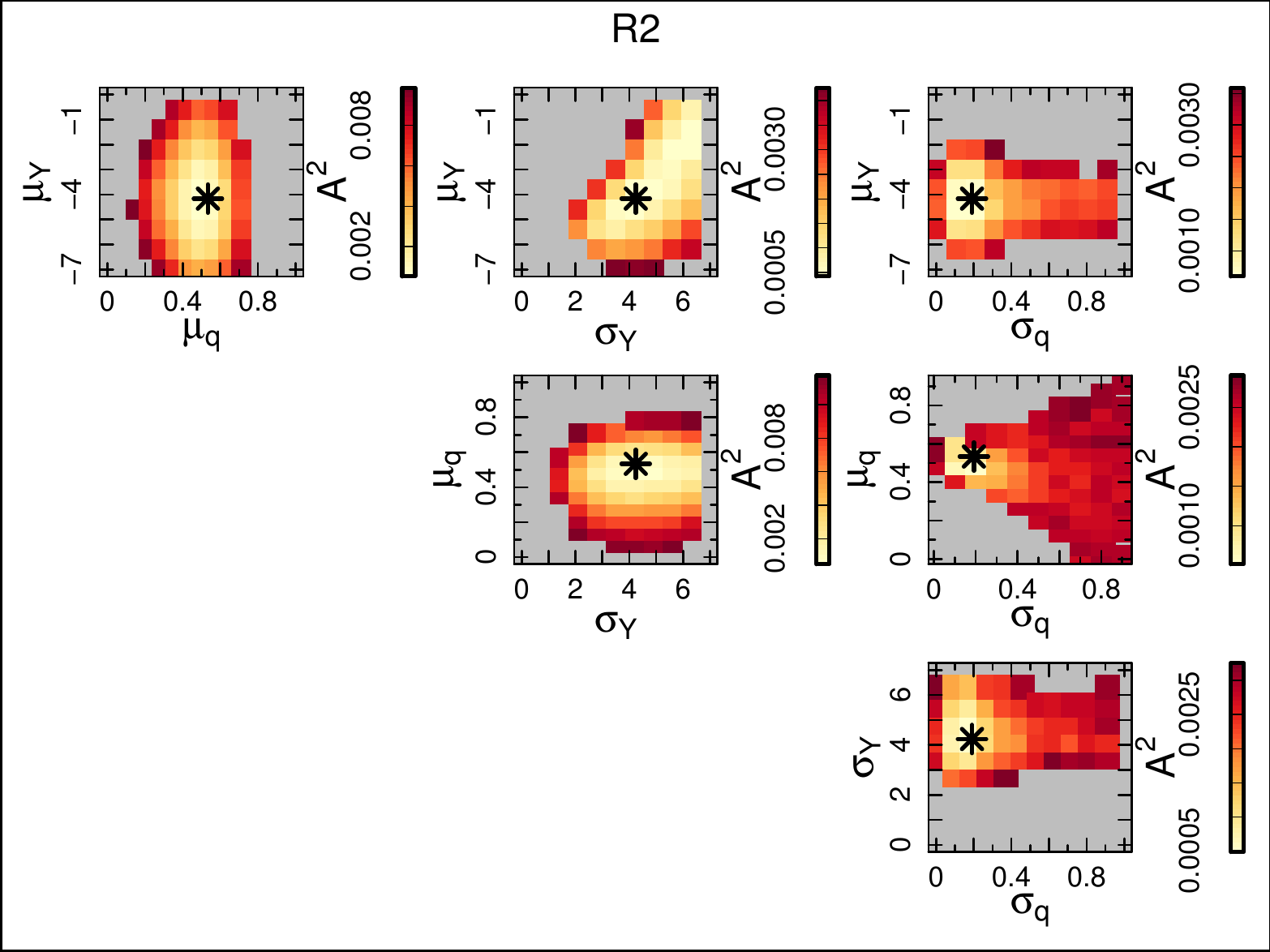}\\
\includegraphics[width=80mm]{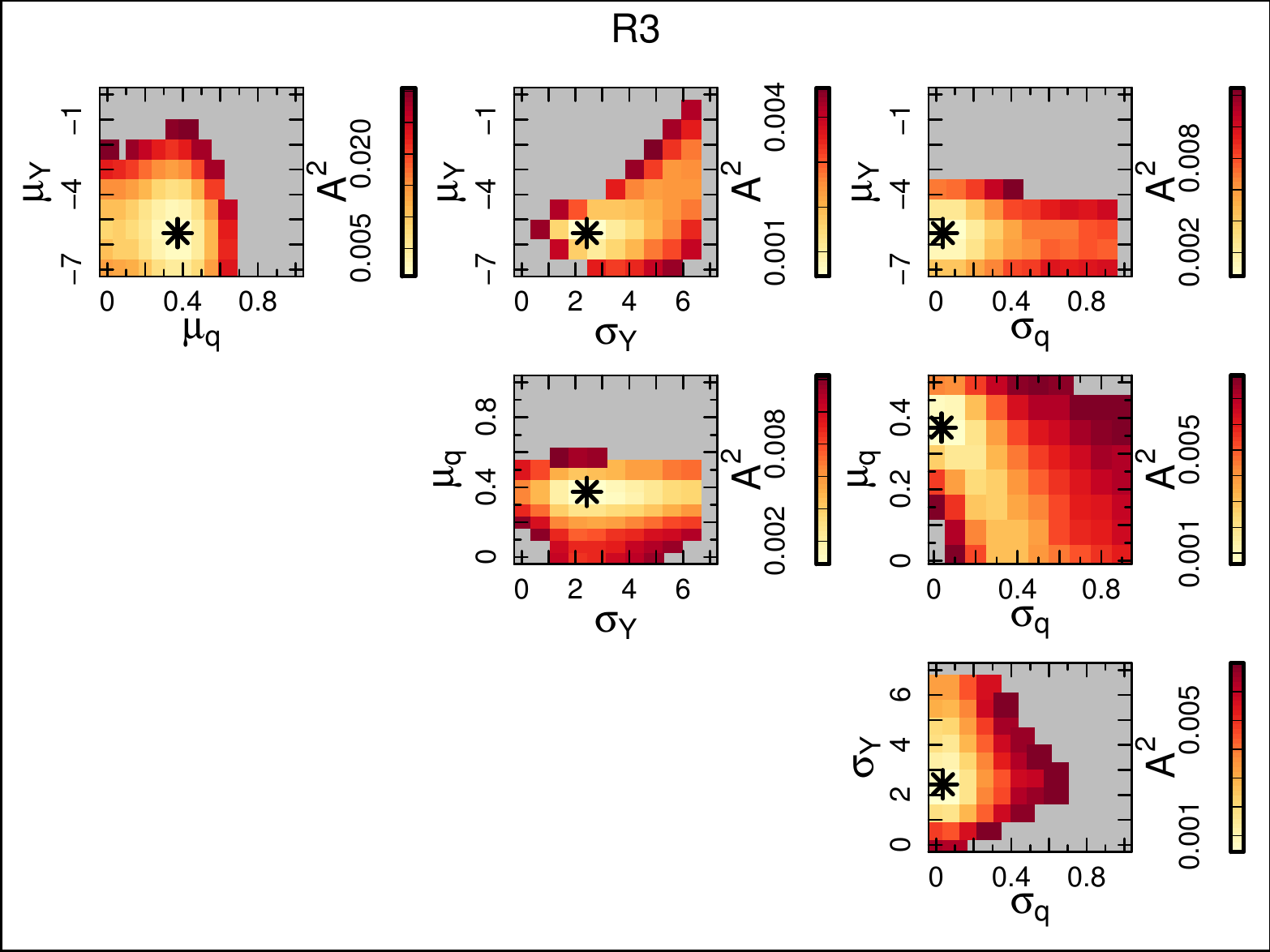}\includegraphics[width=80mm]{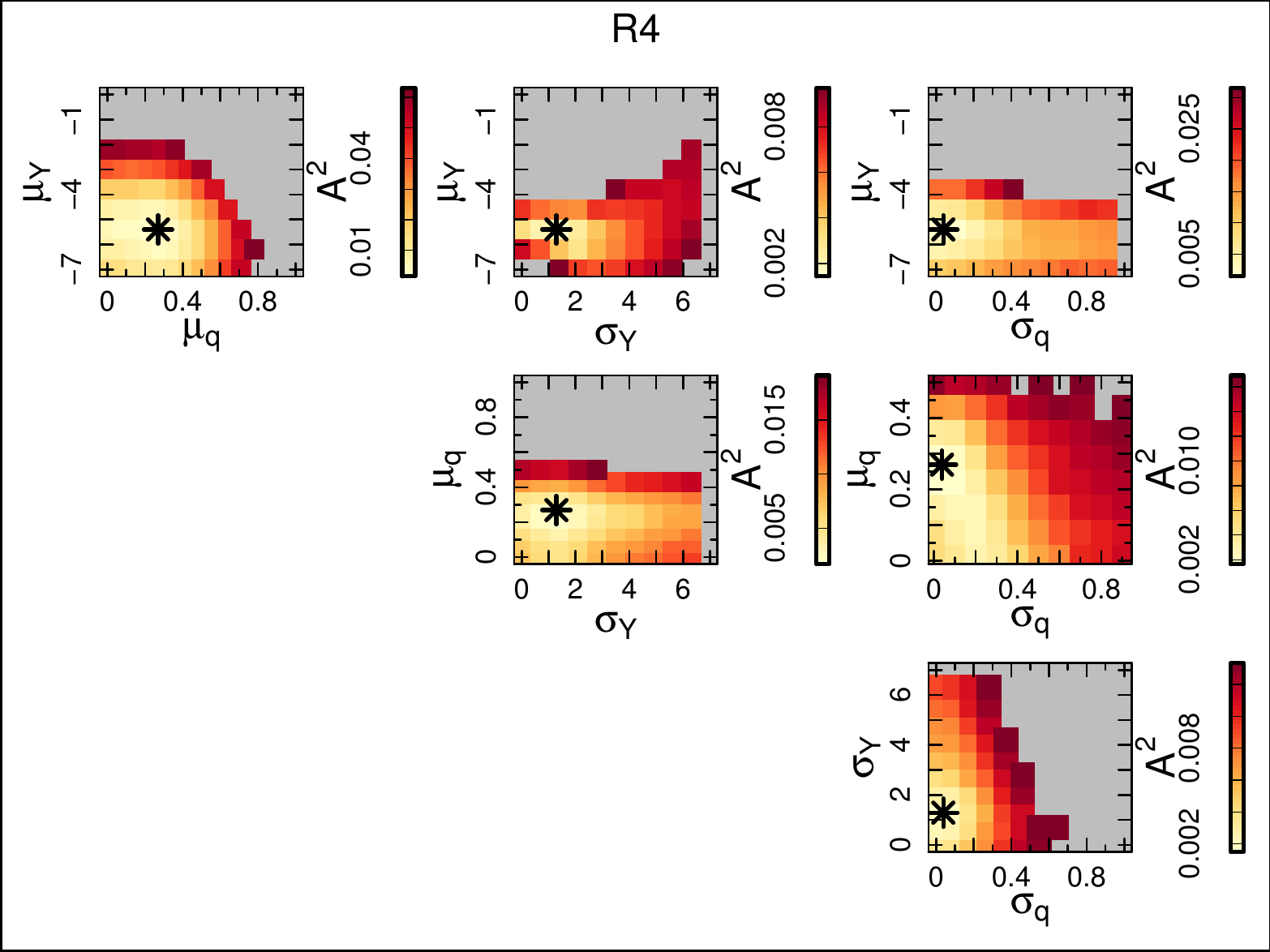}
\caption{\bf Same as Fig. \ref{fig:allA2} for R1 (top left), R2 (top right), R3 (lower left) and R4 (lower right).}\label{fig:allA2}
\end{center}
\end{figure*}

\bibliographystyle{mn2e}
\bibliography{biblio}

\label{lastpage}

\end{document}